\documentclass[11pt]{article}
\usepackage{amsmath,graphicx,graphics,wrapfig,array,epstopdf}
\usepackage{setspace,paralist,lastpage,relsize,verbatim, enumitem}
\usepackage{amssymb,psfrag}
\usepackage{textcomp}
\usepackage{multirow}
\usepackage{hhline}
\usepackage{cite}
\usepackage{media9}
\usepackage{hyperref}
\usepackage{mathrsfs}
\usepackage[usenames,dvipsnames]{color}
\usepackage{framed}
\usepackage{multirow}

\usepackage{float}

\usepackage{tikz}
\newcommand*\circled[1]{\tikz[baseline=(char.base)]{
\node[shape=circle,draw,inner sep=2pt] (char) {#1};}}

\newtheorem{remark}{Remark}

\usepackage{geometry}

\usepackage{breqn}

\DeclareMathOperator{\atantwo}{atan2}

\newcommand{\pd}[2]{\frac{\partial #1}{\partial #2}}

\title{Feedback-Induced Flutter Instability\\ of a Flexible Beam in Fluid Flow}
\vspace{0.5in}

\author{
\\
\\
Sanders Aspelund
$\quad$ Ranjan Mukherjee\\
Department of Mechanical Engineering\\
Michigan State University\\
East Lansing, MI 48824\\
\\
\\
Aren Hellum\\
Vehicle Dynamics and Signature Control\\
Naval Undersea Warfare Center\\
Newport, RI 02841\\
}

\date{}

\begin{document}

\vspace{0.5in}

\maketitle

\vspace{3.7\baselineskip}

\begin{centering}

\begin{abstract}
A pinned-free beam in axial fluid flow, subjected to feedback-based actuation at the pinned end, is investigated. The actuation may be a moment or a prescribed angle and it is proportional to the state (curvature, slope, or displacement) of the beam at some point along its length. All equations and boundary condition terms are non-dimensionalized and the stability of the system is studied over a range of external flow velocity and sensing location. For each combination of flow velocity and sensing location, the critical gain (positive or negative) for the onset of flutter is determined. This process, which is repeated for each combination of actuation and sensing modes, reveals that the closed-loop system exhibits a rich set of stability transitions, each associated with a traveling waveform in the flexible beam at the onset of flutter. With the intent of exploring the use of flexible fluttering beams for underwater propulsion, the efficiency of these waveforms is computed using slender-body theory. Additional insights into the efficiency of the waveforms are obtained through considerations of the smoothness of the traveling waveforms.
\end{abstract}

\end{centering}
\vspace{0.2in}

\newpage

\section{Introduction}\label{sec1}
Elastic structures can lose stability through divergence or flutter \cite{bolotin1963nonconservative, Ziegler68}. Unlike divergence, which is associated with a zero critical frequency, instability through flutter results in oscillations due to a non-zero critical frequency. Beyond the critical stability point, the amplitude of these oscillations grows and often results in limit cycle behavior due to the pronounced effect of system nonlinearities, for example \cite{kounadisParadoxDestabilizingEffect1992, luongoDestabilizingEffectDamping2014, zamaniAsymmetricPostflutterOscillations2015}. The sustained limit cycle oscillations can be undesirable, such as in the diverse cases of aircraft wings \cite{livneAircraftActiveFlutter2018} and blood vessels \cite{grotbergBiofluidMechanicsFlexible2004}, or desirable, such as in the case where it is exploited for underwater propulsion \cite{hellumFlutterInstabilityFluidconveying2011, hellumModelingSimulationDynamics2013}. Irrespective of the application and the underlying objective, flutter instability investigations understandably continue to remain a fertile area of research.\

Flutter instability typically\footnote{Recently, it has been shown that elastic systems subjected to nonholonomic constraints \cite{cazzolliFlutterInstabilityZiegler2020, cazzolliNonholonomicConstraintsInducing2020} can exhibit flutter instability in the presence of conservative loads} occurs due to non-conservative loading \cite{bolotin1963nonconservative, Ziegler68}. Both discrete and continuous systems have been investigated with non-conservative loads generated by a follower force or fluid flow. Some of the early work on discrete systems investigated the double pendulum subjected to a follower force\footnote{This system is often referred to as Ziegler's double pendulum} \cite{ziegler1952stability, herrmannStabilityElasticSystems1964} and articulated pipes conveying fluid\cite{benjaminDynamicsSystemArticulated1962}. Similarly, for continuous systems, the stability of a cantilevered column subjected to a follower force\footnote{This system is often referred to as Beck's column} \cite{beckKnicklastEinseitigEingespannten1952} and fluid-conveying pipes with different boundary conditions \cite{paidoussisPipesConveyingFluid1993} were investigated. A follower force is often viewed as a theoretical construct \cite{elishakoffControversyAssociatedSoCalled2005}, but it can be generated by rocket thrust \cite{sugiyamaFlutterCantileveredColumn1995}, friction \cite{bigoniFlutterDivergenceInstability2018}, and electrostatics \cite{singhMEMSImplementationAxial2005}, for example; a review of the literature on follower forces can be found in \cite{elishakoffControversyAssociatedSoCalled2005, sugiyamaDynamicStabilityColumns2019}. Fluid flow can be internal, external, or both, and typically generates a follower end load along with Coriolis damping; a comprehensive treatise of elastic structures in axial flow can be found in \cite{paidoussisFluidStructureInteractionsSlender2013, paidoussisFluidStructureInteractionsVolume2003}.\

Non-conservative loading can be generated in the absence of a follower force or fluid flow. It has been shown that a dynamic moment acting at the free end of a cantilevered beam, proportional to the slope or curvature of the beam at some point along its length, is non-conservative in nature \cite{abdullatifDivergenceFlutterInstabilities2019} and can result in flutter. It is important to note that this form of non-conservative loading is based on feedback\footnote{The notion of feedback is implicitly present when a follower force is a theoretical construct but it requires sensing only the slope of the free end of the beam. There is no feedback involved when a follower force is implemented in hardware, \cite{bigoniFlutterDivergenceInstability2018}, for example; the direction of the force changes automatically based on the motion of the structure.} where  the measurement can be taken from any point along the length of the beam. In this paper a pinned-free beam in axial fluid flow is actuated at its pinned end; the actuation is based on feedback from sensing the state of the beam at an arbitrary point along its length. Two key differences of this work from \cite{abdullatifDivergenceFlutterInstabilities2019} are that the beam is immersed in fluid flow and the actuation is at the leading edge of the beam. The beam is immersed in fluid flow with the objective of developing a flexible propulsor; the actuator was placed at the leading edge since an actuator (heavy mass) at the trailing edge would impede the generation of traveling waves that are necessary for propulsion.

This work is motivated by our earlier work \cite{hellumFlutterInstabilityFluidconveying2011, hellumModelingSimulationDynamics2013} where internal flow was used to induce flutter and is an extension of the preliminary results presented in \cite{abdullatifCriticalStabilityHinged2019}. Here we consider the possibility of three modes of sensing (curvature, slope, and displacement) and two modes of actuation (moment and prescribed angle). The problem is formally stated in Section \ref{sec2} and the method of solution is given in Section \ref{sec3}. The procedure for determining flutter instability is outlined in Section \ref{sec4} and six illustrative cases are examined in Section \ref{sec5}. These results show the rich set of behaviours that can be generated with three modes of sensing and two modes of actuation. The waveforms produced and their propulsive characteristics are examined in Section \ref{sec6} with the expectation that they will spur further investigation of feedback-induced flutter for propulsion. Concluding remarks are presented in Section \ref{sec7}.

\section{Problem Formulation}\label{sec2}

\begin{figure}[h!]
\centering
\psfrag{A}[][]{\small{$y$}}
\psfrag{B}[][]{\small{$x$}}
\psfrag{C}[][]{\small{$\widehat x$}}
\psfrag{D}[][]{\small{$U_{\rm e}$}}
\psfrag{E}[][]{\small{point of sensing}}
\psfrag{F}[][]{\small{$L$}}
\psfrag{H}[][]{\small{$w$}}
\includegraphics[width=0.64\hsize]{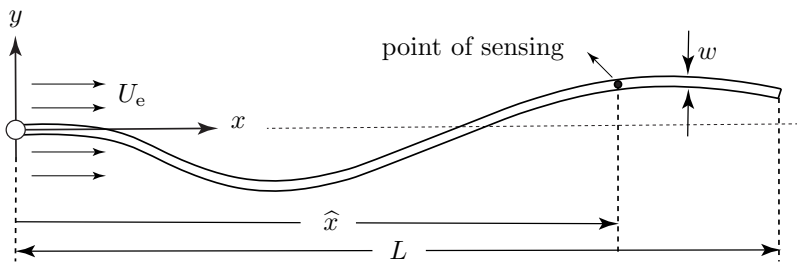}
\caption{A flexible beam, connected at one end to a fixed point by a revolute joint and free at the other end, is immersed in a fluid flowing with constant velocity $U_{\rm e}$.}
\label{Fig1}
\end{figure}

Consider the fluid-immersed flexible beam in Fig.\ref{Fig1}. The beam has length $L$, a rectangular cross-section with width $w$ and height $h$, mass per unit length $m_{\rm b}$, and Young's modulus of elasticity $E$. The upstream end of the beam is connected to a fixed point by a revolute joint, which is actively controlled; the downstream end of the beam is free. The fluid is inviscid and flows with constant velocity $U_{\rm e}$. The equation of motion of the beam, ignoring gravitational, viscous, pressurization and tensile effects, are as follows \cite{paidoussisFluidStructureInteractionsVolume2003, hellumFlutterInstabilityFluidconveying2011}:
\begin{equation}
EI \ \dfrac{\partial^4 y(x, t)}{\partial x^4} + m_e U_{\rm e}^2 \ \dfrac{\partial^2 y(x, t)}{\partial x^2} + 2 m_e U_{\rm e} \ \dfrac{\partial^2 y(x, t)}{\partial x \partial t} +(m_{\rm e} + m_{\rm b}) \ \dfrac{\partial^2 y(x, t)}{\partial t^2}= 0
\label{eq1}
\end{equation}

\noindent where $y(x,t)$ is the displacement of the beam, $I = (h w^3/12)$ is the area moment of inertia of the beam, and $m_e$ is the mass per unit length of the external fluid. The mass per unit length of the external fluid is approximated as the mass of water within the cylinder of unit length circumscribing the beam cross-section \cite{brennenReviewAddedMass1982}.

The displacement boundary condition of the pinned end of the beam and the shear and moment boundary conditions of the free end of the beam are given as
\begin{equation} \label{eq2}
y(0,t) = EI\pd{^2y(L,t)}{x^2} = EI\pd{^3y(L,t)}{x^3} = 0
\end{equation} 

\noindent We assume that an actuator located at the revolute joint can either apply a bending moment or impose an angle condition on the beam at $x=0$. Furthermore, the bending moment or the imposed angle can be based on feedback: proportional to the curvature, slope, or displacement of the beam at some point along its length $x=\widehat{x}$. Therefore, the final boundary condition at $x=0$ can take one of six forms depending on the two modes of actuation and three modes of feedback:
\begin{subequations} \label{eq3}
\begin{align}
EI\pd{^2y(0,t)}{x^2} &= C_{\rm m,c}\, \pd{^2y(\widehat x,t)}{x^2} & &\text{moment $\propto$ curvature} \label{eq3a}\\
EI\pd{^2y(0,t)}{x^2} &= C_{\rm m,s}\, \pd{y(\widehat x,t)}{x} & &\text{moment $\propto$ slope}  \label{eq3b}\\
EI\pd{^2y(0,t)}{x^2} &= C_{\rm m,d}\, y(\widehat x,t) & &\text{moment $\propto$ displacement}  \label{eq3c}\\
\pd{y(0,t)}{x} &= C_{\rm a,c}\, \pd{^2y(\widehat x,t)}{x^2} & &\text{angle $\propto$ curvature}  \label{eq3d}\\
\pd{y(0,t)}{x} &= C_{\rm a,s}\, \pd{y(\widehat x,t)}{x} & &\text{angle $\propto$ slope}  \label{eq3e}\\
\pd{y(0,t)}{x} &= C_{\rm a,d}\, y(\widehat x,t) & &\text{angle $\propto$ displacement}  \label{eq3f}
\end{align}
\end{subequations}
\noindent where $C_{\rm m,c}$, $C_{\rm m,s}$, $C_{\rm m,d}$, $C_{\rm a,c}$, $C_{\rm a,s}$ and $C_{\rm a,d}$ are feedback gains of appropriate dimensions.

\begin{remark}
For the purpose of theoretical development, it is assumed that the curvature, slope, or displacement of the beam at an arbitrary point along its length can be measured directly using sensors or estimated from sensor measurements.
\end{remark}

\begin{remark}
The actuator at the revolute joint can directly apply a moment proportional to the curvature, slope, or displacement at some point along the length of the beam. To impose an angle condition, whereby the revolute joint angle is proportional to the curvature, slope, or displacement, a feedback controller must be designed to drive the actuator to track the desired angle. 
\end{remark}

We introduce the following change of variables:
\begin{equation*}
    v = \frac{y}{L}, \qquad u = \frac{x}{L}, \qquad \gamma = \frac{\widehat x}{L}, \qquad u_{\rm e} = U_{\rm e}L \sqrt{\frac{m_{\rm e}}{EI}}, \qquad \tau = t \sqrt{\frac{EI}{(m_{\rm e}+m_{\rm b})L^4}}
\end{equation*}

\noindent to obtain the non-dimensional equation of motion of the beam
\begin{equation}\label{eq4}
    \pd{^4v}{u^4} + u_{\rm e}^2\, \pd{^2v}{u^2} + 2 \sqrt{\beta}\, u_{\rm e}\, \pd{^2v}{u \partial \tau} + \pd{^2v}{\tau^2} = 0
\end{equation}

\noindent where $\beta$ is the mass fraction:
\begin{equation*}\label{eq:beta}
     \beta = \frac{m_{\rm e}}{m_{\rm e}+m_{\rm b}}
\end{equation*}

\noindent From \eqref{eq2}, the non-dimensional displacement boundary condition of the pinned end of the beam and the non-dimensional natural boundary conditions of the free end of the beam are given as
\begin{equation} \label{eq5}
v(0,\tau) = \pd{^2v(1,\tau)}{u^2} = \pd{^3v(1,\tau)}{u^3} = 0.
\end{equation}

\noindent From \eqref{eq3}, the actuator-imposed boundary condition at the revolute joint takes one of the following six forms:
\begin{subequations} \label{eq6}
\begin{align}
\pd{^2v(0,\tau)}{u^2} &= c_{\rm m,c}\, \pd{^2v(\gamma,\tau)}{u^2} & &\text{moment $\propto$ curvature} \label{eq6a}\\
\pd{^2v(0,\tau)}{u^2} &= c_{\rm m,s}\, \pd{v(\gamma,\tau)}{u} & &\text{moment $\propto$ slope}  \label{eq6b}\\
\pd{^2v(0,\tau)}{u^2} &= c_{\rm m,d}\, v(\gamma,\tau) & &\text{moment $\propto$ displacement}  \label{eq6c}\\
\pd{v(0,\tau)}{u} &= c_{\rm a,c}\, \pd{^2v(\gamma,\tau)}{u^2} & &\text{angle $\propto$ curvature}  \label{eq6d}\\
\pd{v(0,\tau)}{u} &= c_{\rm a,s}\, \pd{v(\gamma,\tau)}{u} & &\text{angle $\propto$ slope}  \label{eq6e}\\
\pd{v(0,\tau)}{u} &= c_{\rm a,d}\, v(\gamma,\tau) & &\text{angle $\propto$ displacement}  \label{eq6f}
\end{align}
\end{subequations}

\noindent where the non-dimensional feedback gains in \eqref{eq6} are related to their dimensional counterparts by the relations
\begin{equation*}\label{nondimparametervalues}
c_{\rm m,c} = \frac{C_{\rm m,c}}{E I}, \quad 
c_{\rm m,s} = \frac{L C_{\rm m,s}}{E I}, \quad 
c_{\rm m,d} = \frac{L^2 C_{\rm m,d}}{E I}, \quad
c_{\rm a,c} = \frac{C_{\rm a,c}}{L}, \quad 
c_{\rm a,s} = C_{\rm a,s}, \quad 
c_{\rm a,d} = L C_{\rm a,d}
\end{equation*}

\section{Method of Solution}\label{sec3}

To solve \eqref{eq4} for the boundary conditions in \eqref{eq5} and \eqref{eq6}, we followed the procedure introduced in \cite{paidoussisFluidStructureInteractionsVolume2003} and used in \cite{hellumFlutterInstabilityFluidconveying2011}. In particular, we assume the following separable form for $v(u,\tau)$:
\begin{equation}
v(u,\tau)=f(u)e^{i \Omega \tau}
\label{eq7}
\end{equation}
\noindent where $\Omega$ is the non-dimensional frequency of oscillation. Substitution of \eqref{eq7} into \eqref{eq4} and \eqref{eq5} yields
 \begin{equation}
f^{\prime \prime \prime \prime}(u)+ u_{\rm e}^2 f^{\prime \prime}(u)+2 u_{\rm e} \sqrt{\beta}\, i \Omega f^{\prime}(u) - \Omega^2 f(u)=0
\label{eq8}
\end{equation}
\begin{equation}\label{eq9}
f(0)  = f^{\prime \prime}(1) = f^{\prime \prime\prime}(1) = 0
\end{equation}

\noindent while substitution of \eqref{eq7} into \eqref{eq6} yields
\begin{subequations} \label{eq10}
\begin{align}
f^{\prime \prime}(0) &= c_{\rm m,c}\, f^{\prime \prime}(\gamma) & &\text{moment $\propto$ curvature} \label{eq10a}\\
f^{\prime \prime}(0) &= c_{\rm m,s}\, f^{\prime}(\gamma) & &\text{moment $\propto$ slope}  \label{eq10b}\\
f^{\prime \prime}(0) &= c_{\rm m,d}\, f(\gamma) & &\text{moment $\propto$ displacement}  \label{eq10c}\\
f^{\prime}(0) &= c_{\rm a,c}\, f^{\prime \prime}(\gamma) & &\text{angle $\propto$ curvature}  \label{eq10d}\\
f^{\prime}(0) &= c_{\rm a,s}\, f^{\prime}(\gamma) & &\text{angle $\propto$ slope}  \label{eq10e}\\
f^{\prime}(0) &= c_{\rm a,d}\, f(\gamma) & &\text{angle $\propto$ displacement}  \label{eq10f}
\end{align}
\end{subequations}

\noindent Since \eqref{eq8} is an ordinary differential equation with constant coefficients, the solution of $f(u)$ is assumed to be of the form $f(u) = A e^{z u}$; this results in the characteristic equation
\begin{equation}
z^4+ u_{\rm e}^2 z^2+2 u_{\rm e} \sqrt{\beta}\, i \Omega z - \Omega^2 = 0
\label{eq11}
\end{equation}
\noindent For specific values of $u_{\rm e}$ and $\beta$, \eqref{eq11} provides four roots of $z_n$, $n = 1, 2, 3, 4$, which are functions of $\Omega$. The solution of $f(u)$ takes the form 
\begin{equation}
f(u)=A_1 e^{z_1 u} + A_2 e^{z_2 u} + A_3 e^{z_3 u} + A_4 e^{z_4 u}
\label{eq12}
\end{equation}

\noindent Substitution of the boundary conditions in \eqref{eq9} and \eqref{eq10} yields
\begin{equation}
\underbrace{\begin{bmatrix}
1 & 1 & 1 & 1 \\
z_1^2 e^{z_1}  & z_2^2 e^{z_2}  & z_3^2 e^{z_3}  & z_4^2 e^{z_4}  \\
z_1^3 e^{z_1} & z_2^3 e^{z_2}  & z_3^3 e^{z_3}  & z_4^3 e^{z_4} \\
\delta_1 & \delta_2 & \delta_3 & \delta_4
\end{bmatrix}
}_{\mathbb{Z}}
\begin{bmatrix} A_1 \\ A_2 \\ A_3 \\ A_4 \end{bmatrix} =
\begin{bmatrix} 0 \\ 0 \\ 0 \\ 0 \end{bmatrix} 
\label{eq13}
\end{equation}

\noindent where $\delta_n$, $n=1,2,3,4,$ are defined as follows 
\begin{align*}
\delta_n \triangleq \left\{ \begin{array} {l@{\quad:\quad}l}
z_n^2 - c_{\rm m,c}\, z_n^2 e^{z_n \gamma} &\text{moment $\propto$ curvature} \\
z_n^2 - c_{\rm m,s}\, z_n e^{z_n \gamma} &\text{moment $\propto$ slope}  \\
z_n^2 - c_{\rm m,d}\, e^{z_n \gamma} &\text{moment $\propto$ displacement}  \\
z_n - c_{\rm a,c}\, z_n^2 e^{z_n \gamma} &\text{angle $\propto$ curvature}  \\
z_n - c_{\rm a,s}\, z_n e^{z_n \gamma} &\text{angle $\propto$ slope} \\
z_n - c_{\rm a,d}\, e^{z_n \gamma} &\text{angle $\propto$ displacement}  
\end{array} \right.
\end{align*}

\noindent For each of the six modes of actuation and feedback combinations, non-trivial solutions of \eqref{eq13} can be obtained by solving the transcendental equation $\det{(\mathbb{Z})} = 0$. For specific values of $u_{\rm e}$, $\beta$, $\gamma$, and the appropriate feedback gain 
\begin{equation}
c \in \{c_{\rm m,c},\, c_{\rm m,s},\, c_{\rm m,d},\, c_{\rm a,c},\, c_{\rm a,s},\, c_{\rm a,d} \} \label{eq14}
\end{equation}
\noindent the transcendental equation can be solved numerically to get the complex frequencies $\Omega_i$, $i = 1, 2, \cdots$, and the $z_n$ terms, $n=1, 2, 3, 4$, for each $\Omega_i$.\

\section{Investigation of Flutter Instability}\label{sec4}
\subsection{Critical Stability}\label{sec41}

While Section \ref{sec3} provides the frequencies of oscillation, $\Omega_i$, $i = 1, 2, \cdots$, for specific values of $u_{\rm e}$, $\beta$, $\gamma$, and $c$\footnote{The discussion here is general and applies to all six modes of actuation and feedback, \emph{i.e.}, $c$ can be any element of the set in \eqref{eq14}}, we seek to find the critical stability points where the system loses stability through flutter. For a particular $\Omega$ and corresponding $z_n$ terms, $n=1, 2, 3, 4$, the solution of \eqref{eq4}, \eqref{eq5}, and \eqref{eq6} can be obtained by substituting \eqref{eq12} into \eqref{eq7}:
\begin{equation}
v(u, \tau) = \sum_{n=1}^{4} A_n\, e^{z_n u}\, e^{i \Omega \tau} = e^{-{\rm Im}[\Omega]\tau} \sum_{n=1}^{4} A_n\, e^{{\rm Re}[z_n]u}\, e^{i \left\{{\rm Im}[z_n]u + {\rm Re} [\Omega]\tau\right\}}
\label{eq15}
\end{equation}
\noindent where the coefficients $A_n$, $n = 1, 2, 3, 4$, can be obtained from the null space of $\mathbb{Z}$ in \eqref{eq13}. It is clear from \eqref{eq15} that the stability of $v(u,\tau)$ is dependent on the exponential term outside the summation; if ${\rm Im}[\Omega] < 0$, this term is unbounded as $t \rightarrow \infty$. Therefore, the point at which ${\rm Im}[\Omega]$ changes sign from positive to negative represents the onset of flutter instability. The first exponential term inside the summation is bounded because $u$ is bounded; the second exponential term yields periodic motion, because it has an imaginary exponent.\

It should be noted that \eqref{eq15} describes the solution for one non-dimensional frequency $\Omega$. At the flutter instability point, one specific value of $\Omega$, $\Omega = \Omega_{\rm cr}$, satisfies ${\rm Im}[\Omega] = 0$ whereas all other $\Omega$ values satisfy ${\rm Im}[\Omega] > 0$. The frequency $\Omega_{\rm cr}$ is real and is defined as the critical frequency. Since $e^{-{\rm Im}[\Omega]\tau} \rightarrow 0$ as $\tau \rightarrow \infty$ for all $\Omega \neq \Omega_{\rm cr}$, the complete solution at the flutter instability point takes the form
\begin{equation}
v(u, \tau) = \sum_{n=1}^{4} A_n\, e^{{\rm Re}[z_n]u}\, e^{i \left\{{\rm Im}[z_n]u + {\rm Re}[\Omega_{\rm cr}]\tau\right\}}
\label{eq16}
\end{equation}
\noindent Since the imaginary exponent in \eqref{eq16} is a function of both $u$ and $\tau$, the above equation represents a traveling waveform.\

\begin{remark}
In the context of a fluid-immersed slender body, Lighthill \cite{lighthillNoteSwimmingSlender1960} established that a traveling wave can generate positive thrust if the phase velocity of the wave is greater than the fluid velocity. Since the expression in \eqref{eq16} is comprised of four waveforms with different, spatially variable amplitudes and phase velocities, deriving a condition for positive thrust is not straightforward. The propulsive characteristics of the waveform in \eqref{eq16} will be discussed in Section \ref{sec6}.
\end{remark}

\subsection{Numerical Procedure}\label{sec42}
We first determine the natural frequencies $\Omega_i$, $i = 1, 2, \cdots$, for the unforced system, \emph{i.e.}, the system with $u_{\rm e} = 0$ and $c = 0$. The set of natural frequencies are determined separately for the two cases where the moment applied at the revolute joint is zero - pinned boundary condition; and the angle specified at the revolute joint is zero - cantilevered boundary condition. Unlike the cantilevered boundary condition, the pinned boundary condition will include the rigid-body mode; this requires us to include $\Omega_0 = 0$ in the set of natural frequencies for the pinned case. We now introduce the following definition:
\vspace{0.05in}

\noindent \emph{Frequency Band}: The range of frequencies in $(0, \Omega_1]$ is defined as the first frequency band $\Pi_1$. The range of frequencies in $(\Omega_{j-1}, \Omega_j]$ is defined as the $j$-th frequency band $\Pi_j$, $j \geq 2$.
\vspace{0.05in}

To determine the critical stability points, we fix the value of $\beta$ and vary $u_{\rm e}$ and $\gamma$ over some domain. For each point in this domain, we solve for the critical feedback gain $c = c_{\rm cr}$, which causes the system to lose stability through flutter. These points define a surface, which we refer to as the critical stability surface. Each point on the critical stability surface corresponds to a critical frequency $\Omega_{\rm cr}$; these points define a critical frequency surface. The critical stability and frequency surfaces are obtained as follows: For a specific value of $\gamma$, we start with $c = 0$ and $u_{\rm e} = 0.1$. We use the first eleven natural frequencies of the beam, $\Omega_k$, $k = 0, 1, 2, \cdots, 10$, for the pinned case and the first ten natural frequencies of the beam, $\Omega_k$, $k = 1, 2, \cdots, 10$ for the cantilevered case, as the initial guesses to solve for the eigenfrequencies as the magnitude of $c$ is gradually increased. The process is continued until one of the $\Omega_k$'s satisfies the condition ${\rm Im}[\Omega_k] = 0$. This provides the value of $c_{\rm cr}$ and $\Omega_{\rm cr}$ for $u_{\rm e} = 0.1$ and the specific value of $\gamma$; the value of $k$ denotes the mode of flutter instability, which will be formally defined later. The process is repeated by gradually incrementing the value of $u_{\rm e}$ while keeping the value of $\gamma$ fixed; the process is terminated when the value of $c_{\rm cr}$ is uniformly zero\footnote{This signifies that the external flow alone causes the beam to lose stability, much like a flag fluttering in the wind.}. To obtain the critical stability and frequency surfaces, the overall process is repeated on a fine mesh grid for $\gamma$.\

\subsection{Simulation Environment}\label{sec43}

We will investigate flutter instability for $\beta = 0.9822$\footnote{This value of $\beta$ is chosen based on a dimensional example that we will consider later in Section \ref{sec6}.}, 
\begin{align*}
\gamma \in \left\{ \begin{array} {l@{\quad:\quad}l}
\left[0.1, 0.9\right] &\text{feedback based on curvature} \\
\left[0.1, 1.0\right] &\text{feedback based on slope} \\
\left[0.3, 1.0\right] &\text{feedback based on displacement}
\end{array} \right.
\end{align*}
\noindent and
\begin{align*}
u_{\rm e} \in \left\{ \begin{array} {l@{\quad:\quad}l}
\left[0.1, 9.0\right] &\text{actuator applies a bending moment} \\
\left[0.1, 19.0\right] &\text{actuator imposes an angle condition}
\end{array} \right.
\end{align*}

\noindent A different range of $\gamma$ was chosen for each of the feedback modes. Since the beam has zero curvature at its free end, we restrict the upper bound to $0.9$ for curvature feedback; since the beam has zero displacement at the revolute joint, we restrict the lower bound to $0.3$ for displacement feedback. The other bounds were chosen such that the values of the critical feedback gain $c_{\rm cr}$ at the boundaries were not inordinately large compared to those within the bounds.\

The upper bounds on $u_{\rm e}$ were chosen based on the finding that the beam loses stability due to external flow alone at $u_{\rm e} = u_{\rm e, cr} = 8.99$ when the actuator applies a bending moment equal to zero, \emph{i.e.}, pinned boundary condition; and $u_{\rm e} = u_{\rm e, cr} = 18.12$ when the actuator imposes an angle equal to zero, \emph{i.e.}, cantilevered boundary condition. These critical velocities for the pinned and cantilevered boundary conditions are shown in the Argand diagrams in Fig.\ref{Fig2}. The Argand diagrams show the locus of the first few eigenfrequencies as $u_{\rm e}$ is increased from zero; each branch starts at a natural frequency of the system $\Omega_k$.\

The procedure for computing the critical stability points, described in Section \ref{sec42}, can now be better explained with the help of the Argand digrams in Fig.\ref{Fig2}. A specific value of $u_{\rm e} = u_{\rm e}^*$, corresponds to a specific point on each branch of the Argand diagram; note that these points correspond to $c=0$ and therefore the value of $\gamma$ is immaterial. For a specific value of $\gamma = \gamma^*$, increasing the value of $c$ from zero results in eleven (ten) loci of the eigenfrequencies for the moment actuation (angle actuation) case that start on each of the branches of the appropriate Argand diagram at the points corresponding to $u_{\rm e} = u_{\rm e}^*$ and $c=0$. The critical value of $c = c_{\rm cr}$, corresponds to the lowest value of $c$ for which one of the eigenfrequencies satisfy ${\rm Im}[\Omega] = 0$. We now introduce the following definition:
\vspace{0.05in}

\begin{figure}[t!]
\centering
\psfrag{M}[][]{\small{Re($\Omega$)}}
\psfrag{N}[][]{\small{Im($\Omega$)}}
\psfrag{A}[][]{\scriptsize{$\Omega_1$}}
\psfrag{B}[][]{\scriptsize{$\Omega_2$}}
\psfrag{C}[][]{\scriptsize{$\Omega_3$}}
\psfrag{P}[][]{\scriptsize{$\Omega_0$}}
\psfrag{D}[][]{\scriptsize{$k\!=\!1$}}
\psfrag{E}[][]{\scriptsize{$k\!=\!2$}}
\psfrag{F}[][]{\scriptsize{$k\!=\!3$}}
\psfrag{G}[][]{\scriptsize{$k\!=\!4$}}
\psfrag{R}[][]{\scriptsize{$k\!=\!0$}}
\psfrag{H}[][]{\scriptsize{$u_{\rm e, cr}\!=\!8.99$}}
\psfrag{K}[][]{\scriptsize{$u_{\rm e, cr}\!=\!18.12$}}
\psfrag{L}[][]{\scriptsize{$u_{\rm e}\!=\!9$}}
\includegraphics[width=6.0in]{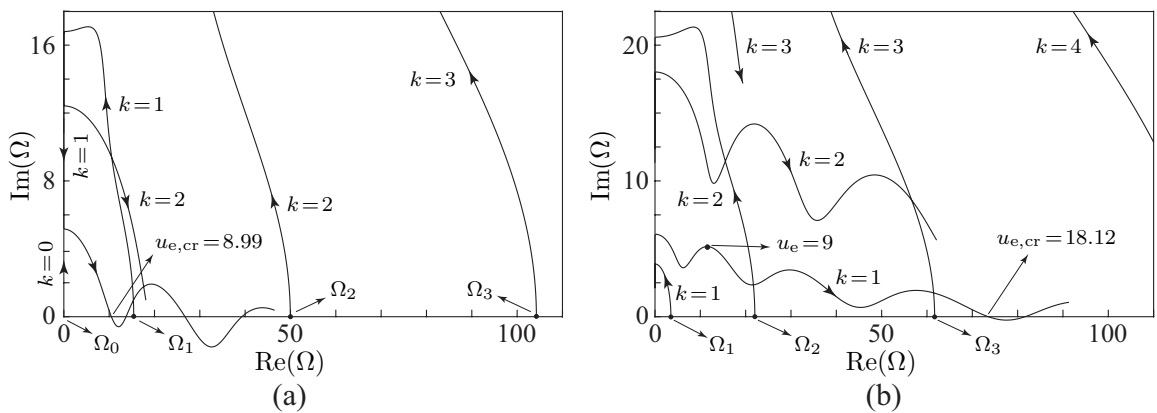}
\caption{\small{Argand diagrams for the beam without feedback for (a) pinned boundary conditions for $u_{\rm e} = [0,15]$ and (b) cantilevered boundary conditions for $u_{\rm e} = [0,20]$. For the pinned case, the locus originating at $\Omega_0$ moves up along the imaginary axis, then downward, and upward again; then, it meets the locus originating at $\Omega_1$ and breaks away into the complex plane resulting in instability for $u_{\rm e, cr} = 8.99$.}}
\label{Fig2}
\end{figure}
\noindent \emph{Single Mode of Flutter Instability}: The system loses stability through the $k$-th mode of flutter if the eigenfrequency satisfying ${\rm Im}[\Omega] = 0$ originated on the $k$-th branch in the Argand diagram of Fig.\ref{Fig2}.
\vspace{0.05in}

\noindent The above definition implicitly assumes that the loci of the eigenfrequencies do not intersect each other prior to satisfying the condition ${\rm Im}[\Omega] = 0$. To account for the possibility of intersection of loci, we introduce the following definition:
\vspace{0.05in}

\noindent \emph{Coupled Mode of Flutter Instability}: The system loses stability through $k_1$-$k_2$ mode of flutter if the eigenfrequency satisfying ${\rm Im}[\Omega] = 0$ can be traced back to the intersection of two loci that originated on the $k_1$-th and $k_2$-th branches of the Argand diagram of Fig.\ref{Fig2}.\
\vspace{0.05in}

Based on the above definitions, in the absence of feedback, stability is lost through the $0$-$1$ mode for the pinned boundary condition of Fig.\ref{Fig2}(a) and through the $1$st mode for the cantilevered boundary condition of Fig.\ref{Fig2}(b).\

\begin{remark}
The coupled mode of flutter is the result of two loci intersecting on the imaginary axis of the Argand diagram. Alternatively, when two loci approach each other in the complex plane but do not intersect before moving away, they exhibit the phenomenon of veering \cite{pierreModeLocalizationEigenvalue1988, abdullatifEffectIntermediateSupport2020}.
\end{remark}

It should be mentioned that both positive and negative values of the feedback gain $c$ can cause the system to lose stability. This means that for each of the three modes of feedback and two modes of actuation, there are two cases to be considered, namely $c > 0$ and $c < 0$. This results in twelve cases of which we present a subset of six illustrative cases which are categorized in Table \ref{Tab1}. It should be mentioned that some of the cases not presented here exhibit divergence mode of instability over a large region of the $\gamma$-$u_{\rm e}$ domain; these include the two cases: moment $\propto$ slope with $c < 0$, and moment $\propto$ displacement with $c < 0$.

\begin{table}[h!]\begin{center}
\caption{Six specific cases chosen for simulation.}\label{Tab1}
\vspace{0.05in}
\begin{tabular}{|l|c|c|c|c|c|c|}
\hline &&&&&&\\[-2.25ex]
Case & \circled{1} & \circled{2}  & \circled{3} &\circled{4} & \circled{5} & \circled{6} \\ &&&&&&\\[-2.25ex]\hline
Actuation &Moment &Angle &Moment &Angle &Moment &Angle \\ \hline
Feedback &Curvature &Curvature &Slope &Slope &Displacement &Displacement \\ \hline
Sign of $c$ & $c<0$ & $c<0$ & $c>0$ & $c<0$ & $c>0$ & $c<0$ \\ \hline
\end{tabular}\end{center}
\end{table}

We complete this section by providing the first eleven (ten) natural frequencies of the beam for the pinned (cantilevered) boundary conditions in Table \ref{Tab2}. These values will be useful when we present our results on the mode of flutter instability and the frequency band in which the system loses stability in the next few subsections. The Argand diagrams in Fig.\ref{Fig2} indicate that, in the absence of feedback, external flow results in $0$-$1$ mode of flutter instability in the first frequency band for the pinned boundary condition, and $1$st mode of flutter instability in the fourth frequency band for the cantilevered boundary condition.\
\begin{table}[h!]\begin{center}
\caption{Natural frequencies of beam for pinned and cantilevered boundary conditions.}\label{Tab2}
\vspace{0.05in}
\begin{tabular}{|l|c|c|c|c|c|c|c|c|c|c|c|}
\hline &&&&&&&&&&&\\[-2.25ex]
&$\Omega_0$ &$\Omega_1$ &$\Omega_2$ &$\Omega_3$ &$\Omega_4$ &$\Omega_5$ &$\Omega_6$ &$\Omega_7$ &$\Omega_8$ &$\Omega_9$ &$\Omega_{10}$\\ &&&&&&&&&&&\\[-2.25ex]\hline
Pinned &0 &15.4 &50.0 &104 &178 &272 &386 &519 &672 &844 &1037 \\ \hline
Cantilevered &- &3.52 &22.0 &61.7 &121 &200 &299 &417 &555 &713 &891 \\ \hline
\end{tabular}\end{center}
\end{table}

\section{Results of Feedback-Induced Instability}\label{sec5}
\subsection{Case 1: Moment $\propto$ Curvature with Negative Feedback Gain}\label{sec51}

\begin{figure}[t!]
    \centering
        \psfrag{A}[l][]{\tiny{$\Pi_1$}}
        \psfrag{B}[l][]{\tiny{$\Pi_2$}}
        \psfrag{C}[l][]{\tiny{$\Pi_3$}}
        \psfrag{D}[l][]{\tiny{$\Pi_4$}}
        \psfrag{E}[l][]{\tiny{$\Pi_5$}}
        \psfrag{F}[l][]{\tiny{$\Pi_6$}}
        \psfrag{G}[l][]{\tiny{$\Pi_7$}}
        \psfrag{H}[l][]{\tiny{$\Pi_8$}}
        \psfrag{J}[l][]{\tiny{$\Pi_9$}}
        \psfrag{K}[l][]{\tiny{$\Pi_{10}$}}
        \psfrag{m}[l][]{\tiny{$0$}}
        \psfrag{a}[l][]{\tiny{$\Omega_1$}}
        \psfrag{b}[l[]{\tiny{$\Omega_2$}}
        \psfrag{c}[l[]{\tiny{$\Omega_3$}}
        \psfrag{d}[l[]{\tiny{$\Omega_4$}}
        \psfrag{e}[l][]{\tiny{$\Omega_5$}}
        \psfrag{f}[l][]{\tiny{$\Omega_6$}}
        \psfrag{g}[l][]{\tiny{$\Omega_7$}}
        \psfrag{h}[l][]{\tiny{$\Omega_8$}}
        \psfrag{j}[l][]{\tiny{$\Omega_9$}}
        \psfrag{k}[l][]{\tiny{$\Omega_{10}$}}
        \psfrag{Q}[][]{\small{$\gamma$}}
        \psfrag{R}[][]{\small{$u_{\rm e}$}}
        \psfrag{S}[][]{\small{$\mid\! c_{\rm cr}\!\mid$}}
        \includegraphics[width=1\textwidth]{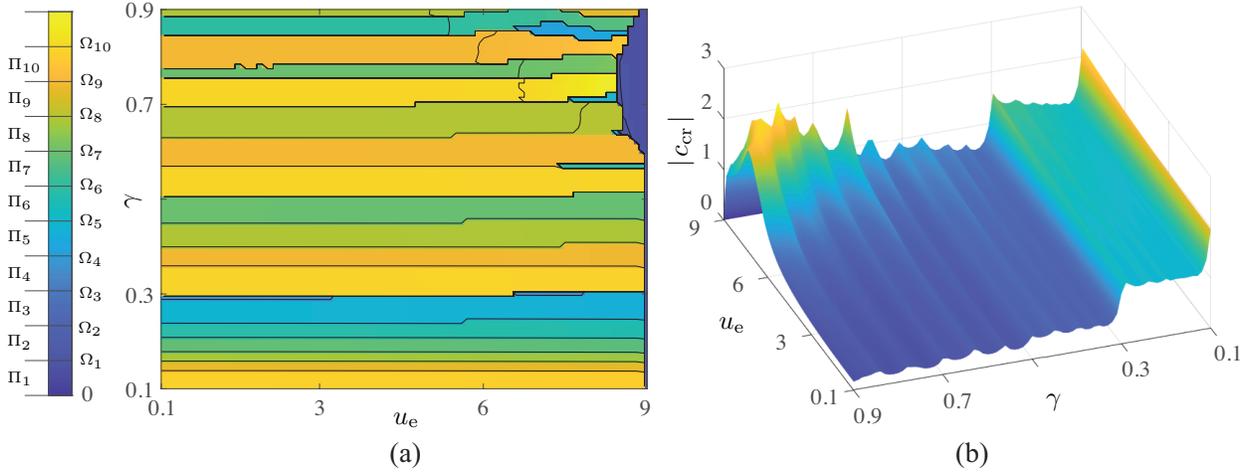}
        \caption{\small{Case 1: (a) Critical frequency surface (b) Critical stability surface. The colorbar pertains only to the critical frequency surface in (a) with the lines demarcating the frequency bands. To better illustrate the topography of the critical stability surface in (b), a suitable perspective view is provided with a color gradient.}}
        \label{Fig3}
\end{figure}

For this case, the moment is proportional to the curvature with feedback gain $c < 0$. For $c=0$, the moment is zero, which signifies the pinned boundary condition. Therefore, the values of the natural frequencies $\Omega_k$, $k = 0, 1, 2, \cdots, 10$, are those of the pinned beam in Table \ref{Tab2}. The critical frequency surface is shown in Fig.\ref{Fig3}(a). The different colors correspond to the frequencies in the color bar shown to the left and the lines demarcate the frequency bands $\Pi_j$, $j=1, 2, \cdots, 10$. Figure \ref{Fig3}(a) shows well-defined striations of constant frequency band for any given $\gamma$ indicating that the critical frequency $\Omega_{\rm cr}$ is highly influenced by the value of $\gamma$. It is evident that for a constant $u_{\rm e}$, a change in the location of sensing (value of $\gamma$) can result in discontinuous changes in $\Omega_{\rm cr}$. In contrast, for a constant $\gamma$, $\Omega_{\rm cr}$ changes gradually as $u_{\rm e}$ changes. In some instances, this results in a change in the frequency band of $\Omega_{\rm cr}$, such as at $u_{\rm e} \approx 6$ and $\gamma \approx 0.8$.\

The critical stability surface is shown in Fig.\ref{Fig3}(b). It exhibits crests and troughs corresponding to the striations in the critical frequency surface in Fig.\ref{Fig3}(a), with the crests corresponding to the boundaries of the striations. The $c_{\rm cr}$ values show an increasing trend as $u_{\rm e}$ increases and drop to zero for $u \geq u_{\rm e, cr} \approx 9$, which corresponds to the point where the pinned beam becomes unstable solely due to the external flow - see Fig.\ref{Fig2}(a). The $c_{\rm cr}$ values are the lowest for low $u_{\rm e}$ and high $\gamma$; they increase moderately with decreasing $\gamma$ and more dramatically with increasing $u_{\rm e}$. As $\gamma$ approaches the pinned end of the beam from $0.3$, the magnitude of $c_{\rm cr}$ increases dramatically, once near $\gamma = 0.3$ and again near $\gamma = 0.1$. As $\gamma$ decreases from $0.3$ to $0.1$, it is noteworthy that the value of $\Omega_{\rm cr}$ in Fig.\ref{Fig3}(a) passes through increasingly higher frequency bands for the entire range of $u_{\rm e}$.\

\subsection{Case 2: Angle $\propto$ Curvature with Negative Feedback Gain}\label{sec52}

\begin{figure}[t!]
    \centering
        \psfrag{A}[l][]{\tiny{$\Pi_1$}}
        \psfrag{B}[l][]{\tiny{$\Pi_2$}}
        \psfrag{C}[l][]{\tiny{$\Pi_3$}}
        \psfrag{D}[l][]{\tiny{$\Pi_4$}}
        \psfrag{E}[l][]{\tiny{$\Pi_5$}}
        \psfrag{F}[l][]{\tiny{$\Pi_6$}}
        \psfrag{G}[l][]{\tiny{$\Pi_7$}}
        \psfrag{H}[l][]{\tiny{$\Pi_8$}}
        \psfrag{J}[l][]{\tiny{$\Pi_9$}}
        \psfrag{K}[l][]{\tiny{$\Pi_{10}$}}
        \psfrag{m}[l][]{\tiny{$0$}}
        \psfrag{a}[l][]{\tiny{$\Omega_1$}}
        \psfrag{b}[l[]{\tiny{$\Omega_2$}}
        \psfrag{c}[l[]{\tiny{$\Omega_3$}}
        \psfrag{d}[l[]{\tiny{$\Omega_4$}}
        \psfrag{e}[l][]{\tiny{$\Omega_5$}}
        \psfrag{f}[l][]{\tiny{$\Omega_6$}}
        \psfrag{g}[l][]{\tiny{$\Omega_7$}}
        \psfrag{h}[l][]{\tiny{$\Omega_8$}}
        \psfrag{j}[l][]{\tiny{$\Omega_9$}}
        \psfrag{k}[l][]{\tiny{$\Omega_{10}$}}
        \psfrag{Q}[][]{\small{$\gamma$}}
        \psfrag{R}[][]{\small{$u_{\rm e}$}}
        \psfrag{S}[][]{\small{$\mid\! c_{\rm cr}\!\mid$}}
        \includegraphics[width=1\textwidth]{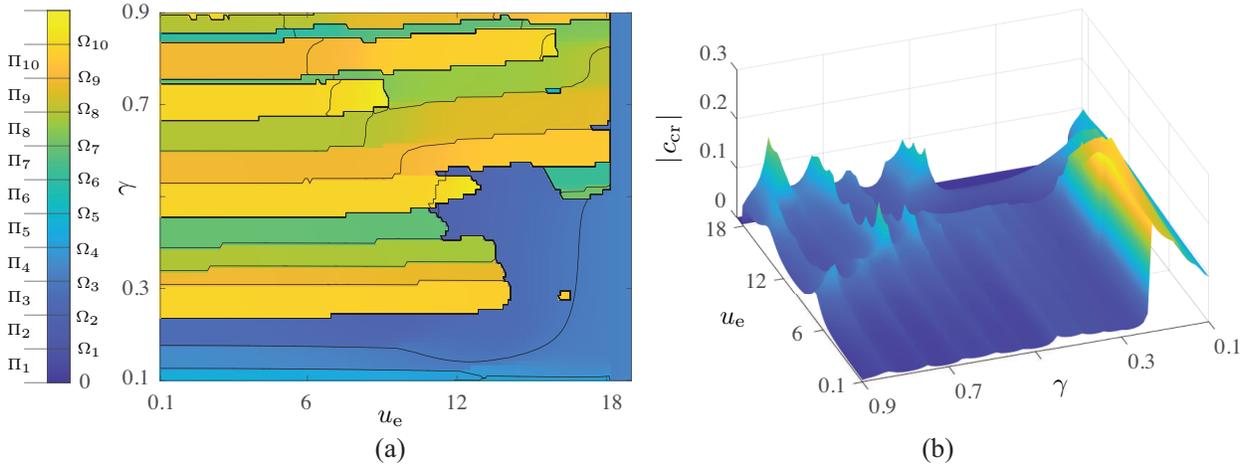}
        \caption{\small{Case 2: (a) Critical frequency surface (b) Critical stability surface. The colorbar pertains only to the critical frequency surface in (a) with the lines demarcating the frequency bands. To better illustrate the topography of the critical stability surface in (b), a suitable perspective view is provided with a color gradient.}}
        \label{Fig4}
\end{figure}

\begin{figure}[b!]
\centering
\psfrag{M}[][]{\small{Re($\Omega$)}}
\psfrag{N}[][]{\small{Im($\Omega$)}}
\psfrag{A}[][]{\scriptsize{$\Omega_1$}}
\psfrag{B}[][]{\scriptsize{$u_{\rm e}^*\!=\!8.87$}}
\psfrag{C}[][]{\scriptsize{$\Omega_3$}}
\psfrag{D}[][]{\scriptsize{$k\!=\!1$}}
\psfrag{E}[][]{\scriptsize{$u_{\rm e}^*\!=\!8.93$}}
\psfrag{F}[][]{\scriptsize{$k\!=\!3$}}
\psfrag{G}[][]{\scriptsize{
\begin{tabular}{l}
$\Omega_{\rm cr}\!=\!29.4$\\ [-0.5ex]
$c_{\rm cr}\!=\!0.21$ \\
\end{tabular}}}
\psfrag{H}[][]{\scriptsize{$k\!=\!8$}}
\psfrag{K}[][]{\scriptsize{$k\!=\!9$}}
\psfrag{J}[][]{\scriptsize{$u_{\rm e}^*\!=\!10.6$}}
\psfrag{L}[][]{\scriptsize{$\gamma\!=\!0.69$}}
\psfrag{P}[][]{\scriptsize{$\gamma\!=\!0.70$}}
\psfrag{S}[][]{\scriptsize{$c_{\rm cr}\!=\!0.15$}}
\psfrag{T}[][]{\scriptsize{$c_{\rm cr}\!=\!0.084$}}
\includegraphics[width=6.0in]{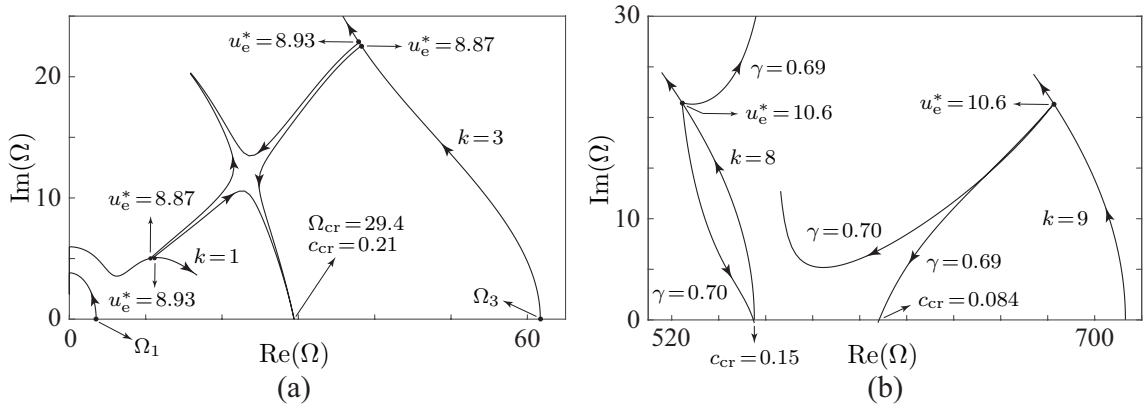}
\caption{\small{Argand diagrams for the beam with feedback for Case 2: (a) a small change in $u_{\rm e}$ results in a change in the mode of flutter instability without significant change in $\Omega_{\rm cr}$, and (b) a small change in $\gamma$ results in a change in both the mode of flutter instability and $\Omega_{\rm cr}$.}}
\label{Fig5}
\end{figure}

For this case, the angle is proportional to the curvature with feedback gain $c < 0$. For $c=0$, the angle is zero, which signifies the cantilevered boundary condition. Therefore, the values of the natural frequencies $\Omega_k$, $k=1, 2, \cdots, 10$, are those of the cantilevered beam in Table \ref{Tab2}. The critical frequency surface is shown in Fig.\ref{Fig4}(a). Similar to Case 1, $\Omega_{\rm cr}$ shows a strong dependence on $\gamma$. However, the striations are wider in general and do not extend for the full range of $u_{\rm e}$. Compared to Case 1, $\Omega_{\rm cr}$ shows a greater dependence on $u_{\rm e}$ and the frequency band of $\Omega_{\rm cr}$ changes with $u_{\rm e}$ for all values of $\gamma > 0.25$. Indeed, for $\gamma \approx 0.8$, $\Omega_{\rm cr}$ smoothly passes through three different frequency bands before exhibiting a sharp drop in frequency at $u_{\rm e} \approx 16$. Unlike Case 1, for $\gamma < 0.58$, several of the striations are terminated ``prematurely" as the system loses stability in low frequency bands for high values of $u_{\rm e}$. The large swath of frequency band $\Pi_3$ originating from $\gamma \in [0.18, 0.24]$ at $u_{\rm e}=0.1$ is not associated with a single mode of flutter. For example, as can be seen from the Argand diagram in Fig.\ref{Fig5}(a), for $\gamma = 0.2$, the mode of flutter changes from $k=3$ for $u_{\rm e} = 8.87$ to $k=1$ for $u_{\rm e} = 8.93$ due to veering \cite{pierreModeLocalizationEigenvalue1988,abdullatifEffectIntermediateSupport2020}; neither $\Omega_{\rm cr}$ nor $c_{\rm cr}$ change significantly.\

The critical stability surface is shown in Fig.\ref{Fig4}(b); it shows crests and troughs along constant $\gamma$ similar to the critical stability surface of Case 1 shown in Fig.\ref{Fig3}(b). Also, the value of $c_{\rm cr}$ becomes zero for $u_{\rm e} \geq u_{\rm e, cr} \approx 18.1$,  which corresponds to the point where the cantilevered beam becomes unstable solely due to the external flow - see Fig.\ref{Fig2}(b). Unlike Case 1, there are additional peaks of $c_{\rm cr}$ in the interior of the domain. The regions around these peaks are associated with changes in the mode and frequencies of flutter instability. For example, for $u_{\rm e} = 10.6$, the mode of flutter changes from $k=9$ for $\gamma = 0.69$ with $c_{\rm cr}=0.084$ to $k=8$ for $\gamma = 0.7$ with $c_{\rm cr}=0.15$; the jump in the $\Omega_{\rm cr}$ and $c_{\rm cr}$ values can be seen from the Argand diagram in Fig.\ref{Fig5}(b).\

\begin{remark}
The magnitude of $c_{\rm cr}$ for Case 2 is $\mathcal{O}(0.1)$; this is an order of magnitude lower than that in Case 1.\
\end{remark}

\subsection{Case 3: Moment $\propto$ Slope with Positive Feedback Gain}\label{sec53}

\begin{figure}[b!]
    \centering
        \psfrag{A}[l][]{\tiny{$\Pi_1$}}
        \psfrag{B}[l][]{\tiny{$\Pi_2$}}
        \psfrag{C}[l][]{\tiny{$\Pi_3$}}
        \psfrag{D}[l][]{\tiny{$\Pi_4$}}
        \psfrag{E}[l][]{\tiny{$\Pi_5$}}
        \psfrag{F}[l][]{\tiny{$\Pi_6$}}
        \psfrag{G}[l][]{\tiny{$\Pi_7$}}
        \psfrag{H}[l][]{\tiny{$\Pi_8$}}
        \psfrag{J}[l][]{\tiny{$\Pi_9$}}
        \psfrag{K}[l][]{\tiny{$\Pi_{10}$}}
        \psfrag{m}[l][]{\tiny{$0$}}
        \psfrag{a}[l][]{\tiny{$\Omega_1$}}
        \psfrag{b}[l[]{\tiny{$\Omega_2$}}
        \psfrag{c}[l[]{\tiny{$\Omega_3$}}
        \psfrag{d}[l[]{\tiny{$\Omega_4$}}
        \psfrag{e}[l][]{\tiny{$\Omega_5$}}
        \psfrag{f}[l][]{\tiny{$\Omega_6$}}
        \psfrag{g}[l][]{\tiny{$\Omega_7$}}
        \psfrag{h}[l][]{\tiny{$\Omega_8$}}
        \psfrag{j}[l][]{\tiny{$\Omega_9$}}
        \psfrag{k}[l][]{\tiny{$\Omega_{10}$}}
        \psfrag{Q}[][]{\small{$\gamma$}}
        \psfrag{R}[][]{\small{$u_{\rm e}$}}
        \psfrag{S}[][]{\small{$\mid\! c_{\rm cr}\!\mid$}}
        \includegraphics[width=1\textwidth]{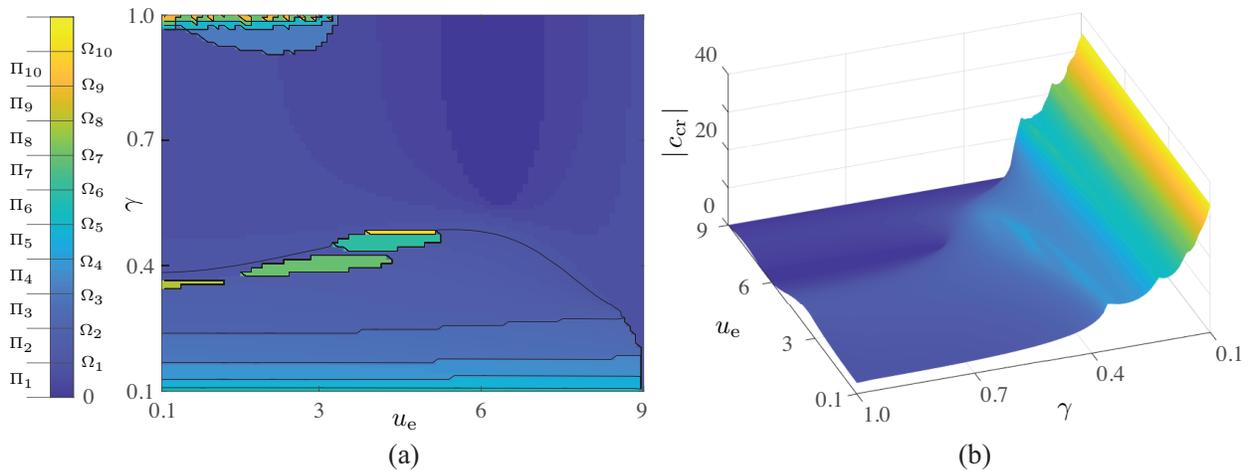}
        \caption{\small{Case 3: (a) Critical frequency surface (b) Critical stability surface. The colorbar pertains only to the critical frequency surface in (a) with the lines demarcating the frequency bands. To better illustrate the topography of the critical stability surface in (b), a suitable perspective view is provided with a color gradient.}}
        \label{Fig6}
\end{figure}

For this case, the moment is proportional to the slope with feedback gain $c > 0$. The values of the natural frequencies $\Omega_k$, $k = 0, 1, 2, \cdots, 10$, are those of the pinned beam in Table \ref{Tab2}. The critical frequency surface is shown in Fig.\ref{Fig6}(a). As the first case to present slope-based feedback, Fig.\ref{Fig6}(a) shows significantly different behavior from the two previous cases. This is in large part due to the behavior of the eigenfrequency originating at $\Omega_0$: this locus is restricted to the imaginary axis for Case 1 involving curvature feedback; for slope-based feedback, it breaks away from the imaginary axis into the complex plane and results in coupled-mode flutter.\

While Fig.\ref{Fig6}(a) shows some distinct striations of frequency bands along constant $\gamma$, these are confined to low values of $\gamma$. In this limited range of $\gamma$, both $\Omega_{\rm cr}$ and $c_{\rm cr}$ decrease as $\gamma$ increases. Contrary to the previous two cases, this trend continues as $\gamma$ increases towards the free end of the beam ($\gamma=1$) resulting in a very large region of $\Pi_1$. Except for small regions where $\gamma \approx 0.4$ and $\gamma \geq 0.95$, stability is lost through flutter in low frequency bands for a large fraction of the $\gamma$-$u_{\rm e}$ domain. In the $\Pi_1$ region, the system loses stability in the 1st mode for low values of $u_{\rm e}$, in the 0-th mode for intermediate values of $u_{\rm e}$, and in the 0-1 coupled mode for high values of $u_{\rm e}$. Stability is lost in the 0-th mode when a pair of loci originating at $\Omega_0$ intersect on the imaginary axis, break away, and cross the real axis at a non-zero frequency. On the other hand, stability is lost in the 0-1 coupled mode when the loci originating at $\Omega_0$ and $\Omega_1$ intersect on the imaginary axis, break away, and cross the real axis. This behavior is similar to what was observed in Fig.\ref{Fig2}(a), albeit for higher values of $u_{\rm e}$ in the absence of feedback.\

The critical stability surface is shown in Fig.\ref{Fig6}(b). Similar to the previous two cases, the value of $c_{\rm cr}$ is low for high values of $\gamma$ and low values of $u_{\rm e}$; the value of $c_{\rm cr}$ increases with decrease in $\gamma$ and increase in $u_{\rm e}$. However, contrary to the previous two cases, the critical stability surface is smooth and contains region in the interior of the domain where $c_{\rm cr} \approx 0$. In particular, for $\gamma \in [0.6, 1.0]$ and $u_{\rm e} \approx 6.3$, the system is marginally stable in the absence of feedback and loses stability with a negligible value of feedback gain $c_{\rm cr}$; the corresponding critical frequency $\Omega_{\rm cr}$ is also small. This region of low $\Omega_{\rm cr}$ and $c_{\rm cr}$ will be discussed further in Section \ref{sec55} with the help of an Argand diagram.\

\begin{remark}
The magnitude of $c_{\rm cr}$ for Case 3 is $\mathcal{O}(10)$; this is an order of magnitude higher than that in Case 1 and two orders of magnitude higher than that in Case 2.\
\end{remark}

\subsection{Case 4: Angle $\propto$ Slope with Negative Feedback Gain}\label{sec54}

\begin{figure}[b!]
    \centering
        \psfrag{A}[l][]{\tiny{$\Pi_1$}}
        \psfrag{B}[l][]{\tiny{$\Pi_2$}}
        \psfrag{C}[l][]{\tiny{$\Pi_3$}}
        \psfrag{D}[l][]{\tiny{$\Pi_4$}}
        \psfrag{E}[l][]{\tiny{$\Pi_5$}}
        \psfrag{F}[l][]{\tiny{$\Pi_6$}}
        \psfrag{G}[l][]{\tiny{$\Pi_7$}}
        \psfrag{H}[l][]{\tiny{$\Pi_8$}}
        \psfrag{J}[l][]{\tiny{$\Pi_9$}}
        \psfrag{K}[l][]{\tiny{$\Pi_{10}$}}
        \psfrag{m}[l][]{\tiny{$0$}}
        \psfrag{a}[l][]{\tiny{$\Omega_1$}}
        \psfrag{b}[l[]{\tiny{$\Omega_2$}}
        \psfrag{c}[l[]{\tiny{$\Omega_3$}}
        \psfrag{d}[l[]{\tiny{$\Omega_4$}}
        \psfrag{e}[l][]{\tiny{$\Omega_5$}}
        \psfrag{f}[l][]{\tiny{$\Omega_6$}}
        \psfrag{g}[l][]{\tiny{$\Omega_7$}}
        \psfrag{h}[l][]{\tiny{$\Omega_8$}}
        \psfrag{j}[l][]{\tiny{$\Omega_9$}}
        \psfrag{k}[l][]{\tiny{$\Omega_{10}$}}
        \psfrag{Q}[][]{\small{$\gamma$}}
        \psfrag{R}[][]{\small{$u_{\rm e}$}}
        \psfrag{S}[][]{\small{$\mid\! c_{\rm cr}\!\mid$}}
        \includegraphics[width=1\textwidth]{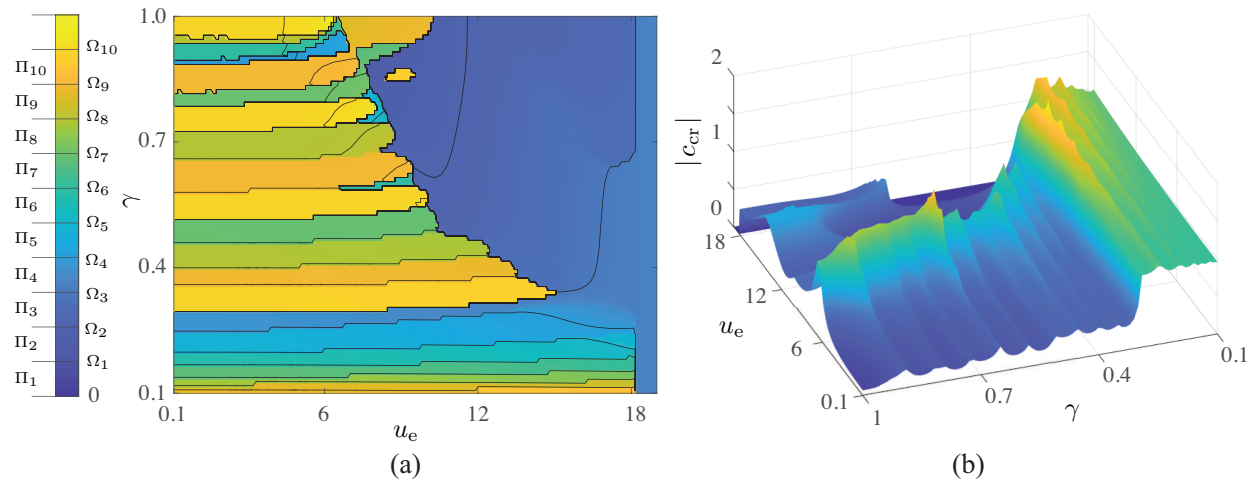}
        \caption{\small{Case 4: (a) Critical frequency surface (b) Critical stability surface. The colorbar pertains only to the critical frequency surface in (a) with the lines demarcating the frequency bands. To better illustrate the topography of the critical stability surface in (b), a suitable perspective view is provided with a color gradient.}}
        \label{Fig7}
\end{figure}

For this case, the angle is proportional to the slope with feedback gain $c > 0$. The values of the natural frequencies $\Omega_k$, $k=1, 2, \cdots, 10$, are those of the cantilevered beam in Table \ref{Tab2}. The critical frequency and critical stability surfaces are shown in Figs.\ref{Fig7}(a) and (b). The striations over the critical frequency surface and crests and troughs over the critical stability surface are quite similar to those observed in Cases 1 and 2. Also similar to these cases, the value of $c_{\rm cr}$ jumps at $\gamma \approx 0.3$ - see Fig.\ref{Fig7}(b). As with all cases discussed so far, $\Omega_{\rm cr}$ increases monotonically as $\gamma$ decreases below $0.3$ - see Fig.\ref{Fig7}(a). Similar to Case 2, there exists a range of $\gamma$, $\gamma \in [0.3, 1.0]$, for which $\Omega_{\rm cr}$ drops abruptly at a specific $u_{\rm e}$, $u_{\rm e} < u_{\rm cr}$; this is exhibited by a sudden change of the frequency bands from high to low in Fig.\ref{Fig7}(a). This sudden drop in $\Omega_{\rm cr}$ is accompanied by a sharp transition in the slope of $c_{\rm cr}$, from rapidly increasing with $u_{\rm e}$ to rapidly decreasing with $u_{\rm e}$ - see Fig.\ref{Fig7}(b).\

The small ``island" of high-frequency band that appears at $u_{\rm e} \approx 9$ and $\gamma \approx 0.85$ in Fig.\ref{Fig7}(a) is associated with the $k=9$ mode of flutter. At $u_{\rm e} \approx 9$, the $k=1$ locus is sufficiently far from the real axis in the Argand diagram in Fig.\ref{Fig2}(b) such that introduction of feedback causes the $k=9$ locus\footnote{The $k=9$ locus is not shown in the Argand diagram in Fig.\ref{Fig2}(b) but shown in Fig.\ref{Fig5}(b) for $u_{\rm e}$ up to at least $10.6$.} to reach the real axis prior to the $k=1$ locus because the $k=9$ locus is very sensitive to $c$ for $\gamma$ values close to $0.85$. When the $k=1$ locus starts sufficiently close to the real axis, introduction of feedback causes the $k=1$ mode to reach the real axis before the other modes - this explains the ``sea" of low-frequency bands in a large fraction of the upper-right domain. The border of this low frequency region shows the effect of the interplay between $\gamma$ and $u_{\rm e}$ on the stability characteristics of the system. In this low-frequency region, the magnitude of $c_{\rm cr}$ exhibits an undulatory behavior as $u_{\rm e}$ increases; this can be attributed to the oscillatory behavior of the $k=1$ locus in the Argand diagram of Fig.\ref{Fig2}(b).\

\begin{remark}
The magnitude of $c_{\rm cr}$ for Case 4 is $\mathcal{O}(1)$, the same magnitude as Case 1 but an order of magnitude higher than that shown in Case 2 and an order of magnitude lower than that shown in Case 3. 
\end{remark}

\subsection{Case 5: Moment $\propto$ Displacement with Positive Feedback Gain}\label{sec55}

\begin{figure}[b!]
    \centering
        \psfrag{A}[l][]{\tiny{$\Pi_1$}}
        \psfrag{B}[l][]{\tiny{$\Pi_2$}}
        \psfrag{C}[l][]{\tiny{$\Pi_3$}}
        \psfrag{D}[l][]{\tiny{$\Pi_4$}}
        \psfrag{E}[l][]{\tiny{$\Pi_5$}}
        \psfrag{F}[l][]{\tiny{$\Pi_6$}}
        \psfrag{G}[l][]{\tiny{$\Pi_7$}}
        \psfrag{H}[l][]{\tiny{$\Pi_8$}}
        \psfrag{J}[l][]{\tiny{$\Pi_9$}}
        \psfrag{K}[l][]{\tiny{$\Pi_{10}$}}
        \psfrag{m}[l][]{\tiny{$0$}}
        \psfrag{a}[l][]{\tiny{$\Omega_1$}}
        \psfrag{b}[l[]{\tiny{$\Omega_2$}}
        \psfrag{c}[l[]{\tiny{$\Omega_3$}}
        \psfrag{d}[l[]{\tiny{$\Omega_4$}}
        \psfrag{e}[l][]{\tiny{$\Omega_5$}}
        \psfrag{f}[l][]{\tiny{$\Omega_6$}}
        \psfrag{g}[l][]{\tiny{$\Omega_7$}}
        \psfrag{h}[l][]{\tiny{$\Omega_8$}}
        \psfrag{j}[l][]{\tiny{$\Omega_9$}}
        \psfrag{k}[l][]{\tiny{$\Omega_{10}$}}
        \psfrag{Q}[][]{\small{$\gamma$}}
        \psfrag{R}[][]{\small{$u_{\rm e}$}}
        \psfrag{S}[][]{\small{$\mid\! c_{\rm cr}\!\mid$}}
        \includegraphics[width=1\textwidth]{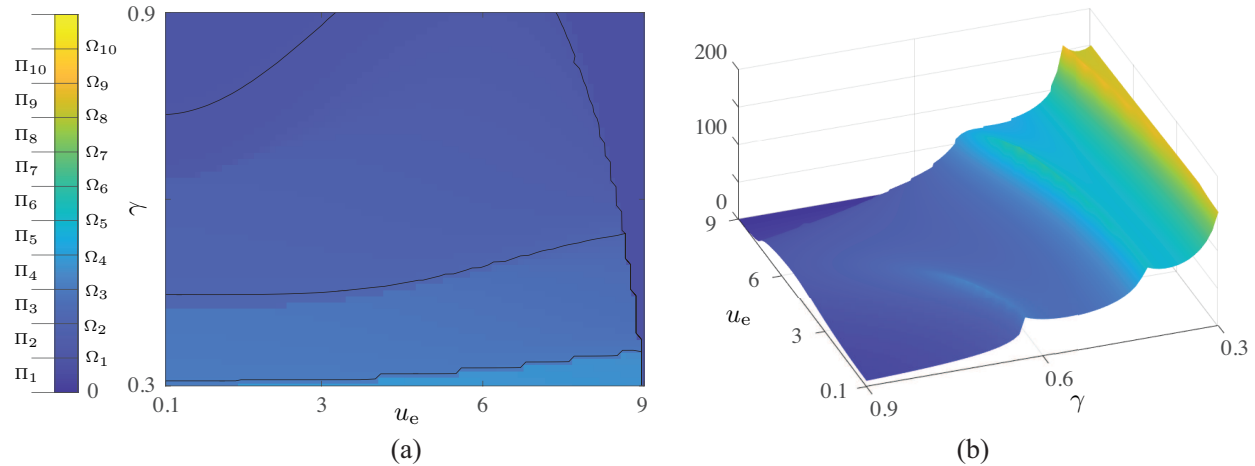}
        \caption{\small{Case 5: (a) Critical frequency surface (b) Critical stability surface. The colorbar pertains only to the critical frequency surface in (a) with the lines demarcating the frequency bands. To better illustrate the topography of the critical stability surface in (b), a suitable perspective view is provided with a color gradient.}}
        \label{Fig8}
\end{figure}

\begin{figure}[t!]
    \centering
    \psfrag{M}[][]{\small{Re($\Omega$)}}
    \psfrag{N}[][]{\small{Im($\Omega$)}}
    \psfrag{A}[][]{\scriptsize{$\Omega_1$}}
    \psfrag{B}[][]{\scriptsize{$\Omega_0$}}
    \psfrag{G}[][]{\scriptsize{$k\!=\!1$}}
    \psfrag{H}[][]{\scriptsize{$k\!=\!0$}}
    \psfrag{K}[][]{\scriptsize{
\begin{tabular}{l}
$\Omega_{\rm cr}\!=\!23.20$\\ [-0.5ex]
$c_{\rm cr}\!=\!29.16$ \\
\end{tabular}}}
\psfrag{E}[][]{\scriptsize{
\begin{tabular}{l}
$\Omega_{\rm cr}\!=\!3.94$\\ [-0.5ex]
$c_{\rm cr}\!=\!2.28$ \\
\end{tabular}}}
\psfrag{C}[][]{\scriptsize{$u_{\rm e}^*\!=\!5$}}
\includegraphics[width=6.0in]{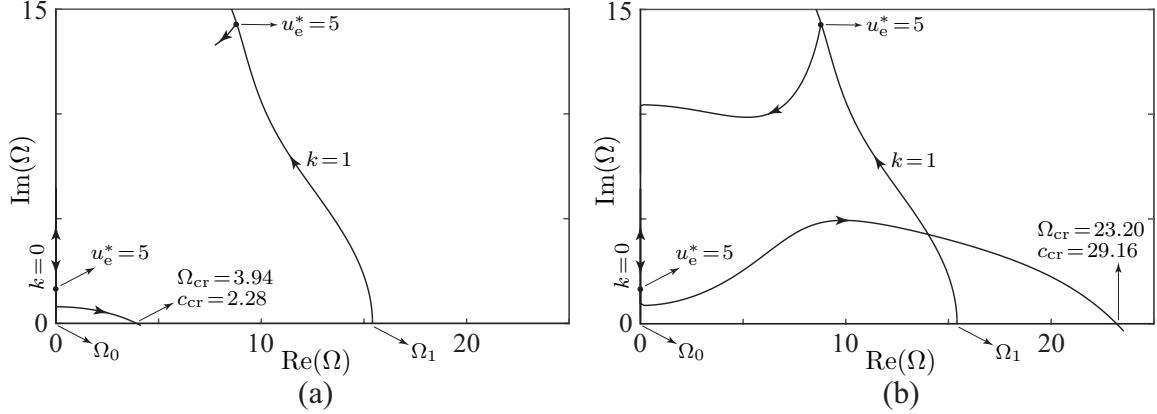}
\caption{\small{Argand diagrams for the beam with slope and displacement feedback. For $u_{\rm e}^*=5.0$ and $\gamma = 0.85$, (a) moment actuation based on slope feedback (Case 3) results in $\Omega_{\rm cr}=3.94$ and $c_{\rm cr}=2.28$; (b) moment actuation based on displacement feedback (Case 5) results in $\Omega_{\rm cr}=23.30$ and $c_{\rm cr}=29.16$.}}
\label{Fig9}
\end{figure}

For this case, the moment is proportional to the displacement with feedback gain $c < 0$. The values of the natural frequencies $\Omega_k$, $k = 0, 1, 2, \cdots, 10$, are those of the pinned beam in Table \ref{Tab2}. The critical frequency and critical stability surfaces are shown in Figs.\ref{Fig8}(a) and (b). Although based on displacement feedback, these plots have many similarities with those of Case 3 in Figs.\ref{Fig6}(a) and (b), which are based on slope feedback. In particular, both the critical frequency and stability surfaces show a nearly monotonic increase with decreasing $\gamma$, with ridges on the critical stability surface demarcating the change in mode of flutter. Also, the frequency bands curve in the direction of increasing $\gamma$ as $u_{\rm e}$ increases; this is distinctly different from the other cases where the narrow frequency bands or striations are largely independent of $u_{\rm e}$. Similar to Case 3, there exists a large region of the $\gamma$-$u_{\rm e}$ domain where the system loses stability in the 1st mode for low values of $u_{\rm e}$, in the 0-th mode for intermediate values of $u_{\rm e}$, and in the 0-1 coupled mode for high values of $u_{\rm e}$. For intermediate values of $u_{\rm e}$ where stability is lost in the 0-th mode, the behavior of the system is however distinctly different from that of Case 3. For slope-based feedback (Case 3), the locus originating at $\Omega_0$ curves towards the real axis immediately after breaking away from the imaginary axis; this results in low values of $\Omega_{\rm cr}$ and $c_{\rm cr}$ - see Fig.\ref{Fig9} (a). For displacement-based feedback (this case), the locus moves away from the real axis and converges on it at a higher value of $\Omega_{\rm cr}$; the associated value of $c_{\rm cr}$ is also higher - see Fig.\ref{Fig9} (b).

\begin{remark}
The magnitude of $c_{\rm cr}$ for Case 5 is $\mathcal{O}(100)$. In comparison to the other two cases of moment actuation, the magnitude of $c_{\rm cr}$ is an order of magnitude higher than Case 3 and two orders of magnitude higher than Case 1.
\end{remark}

\subsection{Case 6: Angle $\propto$ Displacement with Negative Feedback Gain}\label{sec56}

For this case, the angle is proportional to the displacement with feedback gain $c > 0$. The values of the natural frequencies $\Omega_k$, $k = 1, 2, \cdots, 10$, are those of the cantilevered beam in Table \ref{Tab2}. The critical frequency and critical stability surfaces are shown in Figs.\ref{Fig10}(a) and (b); they resemble those of Case 3 in Figs.\ref{Fig6}(a) and (b) and those of Case 5 in Figs.\ref{Fig8}(a) and (b) although the mode of actuation and sign of $c$ are different for both Cases 3 and 5 and the mode of sensing is different for Case 3. The tendency of the critical frequency to increase with increasing $u_{\rm e}$ and decreasing $\gamma$ is particularly noticeable due to the ``curving up" of the frequency bands in Fig.\ref{Fig10}(a); this behavior is present in all cases but is very distinct in Cases 3 and 5. Similar to Case 3, for low $u_{\rm e}$, Fig\ref{Fig10}(a) shows small regions where stability is lost through flutter in high frequency bands. Otherwise, both the critical frequency and stability surfaces resemble Figs.\ref{Fig8}(a) and (b) pertaining to Case 5, increasing towards low $\gamma$ and high $u_{\rm e}$ with ridges on the critical stability surface demarcating the changes in the mode of flutter.\

\begin{remark}
The magnitude of $c_{\rm cr}$ for Case 6 is $\mathcal{O}(10)$. A comparison of all six cases indicate that the magnitude of $c_{\rm cr}$ depends on the modes of actuation and sensing. A change in the mode of actuation from angle to moment increases $c_{\rm cr}$ by one order of magnitude on average. Similarly, changing the mode of sensing from curvature to angle as well as from angle to displacement increases $c_{\rm cr}$ by one order of magnitude on average - see Table \ref{Tab3}.

\begin{figure}[t!]
    \centering
        \psfrag{A}[l][]{\tiny{$\Pi_1$}}
        \psfrag{B}[l][]{\tiny{$\Pi_2$}}
        \psfrag{C}[l][]{\tiny{$\Pi_3$}}
        \psfrag{D}[l][]{\tiny{$\Pi_4$}}
        \psfrag{E}[l][]{\tiny{$\Pi_5$}}
        \psfrag{F}[l][]{\tiny{$\Pi_6$}}
        \psfrag{G}[l][]{\tiny{$\Pi_7$}}
        \psfrag{H}[l][]{\tiny{$\Pi_8$}}
        \psfrag{J}[l][]{\tiny{$\Pi_9$}}
        \psfrag{K}[l][]{\tiny{$\Pi_{10}$}}
        \psfrag{m}[l][]{\tiny{$0$}}
        \psfrag{a}[l][]{\tiny{$\Omega_1$}}
        \psfrag{b}[l[]{\tiny{$\Omega_2$}}
        \psfrag{c}[l[]{\tiny{$\Omega_3$}}
        \psfrag{d}[l[]{\tiny{$\Omega_4$}}
        \psfrag{e}[l][]{\tiny{$\Omega_5$}}
        \psfrag{f}[l][]{\tiny{$\Omega_6$}}
        \psfrag{g}[l][]{\tiny{$\Omega_7$}}
        \psfrag{h}[l][]{\tiny{$\Omega_8$}}
        \psfrag{j}[l][]{\tiny{$\Omega_9$}}
        \psfrag{k}[l][]{\tiny{$\Omega_{10}$}}
        \psfrag{Q}[][]{\small{$\gamma$}}
        \psfrag{R}[][]{\small{$u_{\rm e}$}}
        \psfrag{S}[][]{\small{$\mid\! c_{\rm cr}\!\mid$}}
        \includegraphics[width=1\textwidth]{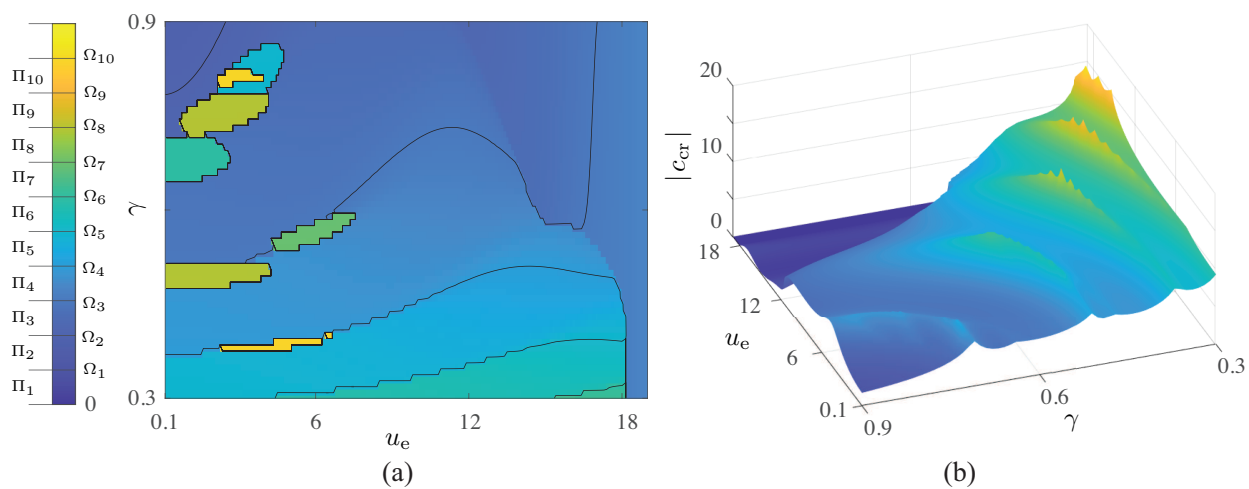}
        \caption{\small{Case 6: (a) Critical frequency surface (b) Critical stability surface. The colorbar pertains only to the critical frequency surface in (a) with the lines demarcating the frequency bands. To better illustrate the topography of the critical stability surface in (b), a suitable perspective view is provided with a color gradient.}}
        \label{Fig10}
\end{figure}

\begin{table}[h!]\begin{center}
\caption{Order of magnitude for critical stability}\label{Tab3}
\vspace{0.05in}
\begin{tabular}{|l|c|c|c|}
\hline 
&Curvature & Slope & Displacement\\
&Sensing & Sensing & Sensing\\
\hline
Moment Actuation &1.0 &10 &100 \\ \hline
Angle Actuation &0.1 &1.0 &10 \\ \hline
\end{tabular}\end{center}
\end{table}
\end{remark}

\section{Application to Underwater Propulsion}\label{sec6}
\subsection{An Underwater Vehicle with a Flexible Propulsor}\label{sec61}

\begin{figure}[b!]
\centering
\psfrag{A}[][]{\small{$y$}}
\psfrag{B}[][]{\small{$x$}}
\psfrag{C}[][]{\small{rigid body}}
\psfrag{D}[][]{\small{$U_{\rm e}$}}
\psfrag{E}[][]{\small{flexible beam}}
\psfrag{F}[][]{\small{$L$}}
\psfrag{H}[][]{\small{active joint}}
\includegraphics[width=0.87\hsize]{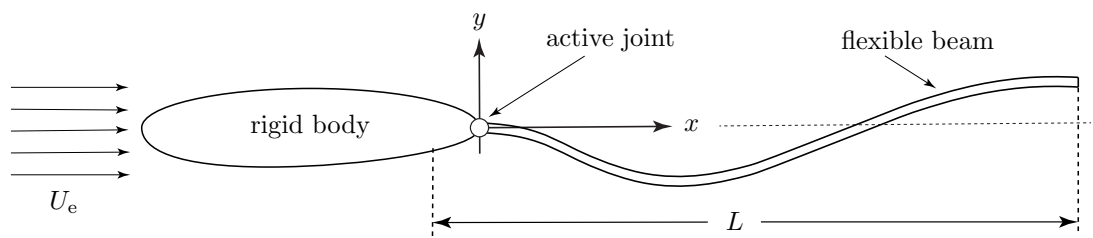}
\caption{A rigid body connected to a tail-like flexible beam by an active revolute joint.}
\label{Fig11}
\end{figure}

All of the analysis presented in Sections \ref{sec2} and \ref{sec3} and the results presented in Section \ref{sec5} can find a potential application in the propulsion of undersea vehicles. To motivate this, we consider a submersible comprised of a rigid body and a tail-like flexible beam, immersed in a quiescent fluid; the rigid body is connected to the flexible beam by an active revolute joint - see Fig.\ref{Fig11}. For the sake of simplicity, we assume that the drag of the submersible is entirely due to the rigid body and the thrust is produced entirely by the flexible tail. The submersible is assumed to move with constant velocity $U_{\rm e}$ in a state of dynamic equilibrium, where thrust and drag forces are equal and opposite. The rigid body is assumed to have negligible rotational motion due to its inertia and consequently it translates with constant velocity $U_{\rm e}$. This is in conformity with the assumptions made in Section \ref{sec2}, namely, the pinned joint is at the origin of an inertial reference frame, and the beam is immersed in a fluid that moves with constant relative velocity $U_{\rm e}$.\

\subsection{Propulsive Characteristics}\label{sec62}
\subsubsection{Thrust, Power, and Efficiency}\label{sec621}

For a ``slender fish", Lighthill \cite{lighthillNoteSwimmingSlender1960} estimated the thrust, power, and efficiency assuming that the fish has neither mass nor area at its leading edge. For the more general case, Hellum \cite{hellumFlutterInstabilityFluidconveying2011} adapted these expressions for computing the nondimensional thrust $\mathbb{F}$ and power $\mathbb{P}$

\begin{subequations}\label{eq17}
\begin{align}
\mathbb{F} &= \frac{\Omega}{4 \pi} \int_{0}^{2\pi/\Omega} \left\{\left[\left(\pd{v}{\tau}\right)^2 - \left(u_{\rm e} \pd{v}{u}\right)^2 \right]_{u=1} - \left[\left(\pd{v}{\tau}\right)^2 - \left(u_{\rm e} \pd{v}{u}\right)^2 \right]_{u=0} \right\}d\tau
\label{eq17a} \\
\mathbb{P} &= \frac{\Omega}{2 \pi} \int_{0}^{2\pi/\Omega} \left\{\left[\pd{v}{\tau}\left(\pd{v}{\tau}+u_{\rm e} \pd{v}{u}\right)\right]_{u=1} - \left[\pd{v}{\tau}\left(\pd{v}{\tau}+u_{\rm e} \pd{v}{u}\right)\right]_{u=0} \right\}d\tau \label{eq17b}
\end{align}
\end{subequations}

\noindent where $\Omega = \Omega_{\rm cr}$ is the non-dimensional frequency of oscillation, and the thrust and power expressions are approximated by their averages over one period of oscillation. The function $v(u,\tau)$ is solved using the procedure outlined in Section \ref{sec3} and has both real and imaginary parts. Since only the real part physically contributes to the thrust and power, $\operatorname{Re}[v]$ is used in place of $v$ in \eqref{eq17} and has the form

\begin{equation}\label{eq18}
\operatorname{Re}[v(u,\tau)] = \sum_{n=1}^{4} \mathrm{e}^{\mathrm{Re}\left[z_{n}\right] u}\left\{\operatorname{Re}\left[A_{n}\right] \cos \left(\operatorname{Im}\left[z_{n}\right] u+\Omega \tau\right)-\operatorname{Im}\left[A_{n}\right] \sin \left(\operatorname{Im}\left[z_{n}\right] u+\Omega \tau\right)\right\}
\end{equation}

\noindent From the thrust and power values, the Froude efficiency \cite{lighthillNoteSwimmingSlender1960} can be calculated as
\begin{equation}\label{eq19}
\eta = \frac{\mathbb{F} u_{\rm e}}{\mathbb{P}}
\end{equation}

Both the thrust and power expressions depend on the amplitude of the waveform of the flexible tail, which depends on the $A_n$ terms in \eqref{eq18}. The $A_n$ terms are obtained from the null-space of $\mathbb{Z}$ in \eqref{eq13}; therefore, the thrust and power will depend on the scaling of the null-space vector. This scaling is arbitrary because we are using a linear model of the system. Ideally, the amplitude of the waveform would be determined by a limit cycle analysis of the nonlinear model, which is outside the scope of this work. Therefore, we focus on the waveform efficiency, which is not dependent on the amplitude of the waveform.\

\subsubsection{Wave Speed and Phase Smoothness}\label{sec622}

The efficiency of the tail-like propulsor, given by \eqref{eq19}, assumes that the thrust generated is positive. This can be verified from the sign of $\mathbb{F}$, computed using \eqref{eq17a} with an arbitrary amplitude of the waveform. For a tail oscillating with a waveform
\begin{equation*}
g(u,\tau) = h(u)\cos(\Omega \tau - k u)
\end{equation*}

\noindent where $k$ is the non-dimensionsal wavenumber, Hellum \cite{hellumFlutterInstabilityFluidconveying2011} provided a simple condition for the tail to generate positive thrust. This condition, which was adapted from Lighthill \cite{lighthillNoteSwimmingSlender1960}, is given as
\begin{equation}\label{eq20}
\frac{\Omega}{k} > u_e\, \beta^{-1/2} \quad \Rightarrow \quad \frac{u_e\, \beta^{-1/2}}{(\Omega/k)} < 1
\end{equation}

\noindent where $(\Omega/k)$ is referred to as the non-dimensional phase velocity. When this condition is met, the efficiency can be alternately computed using the expression
\begin{equation}\label{eq21}
\eta^* = 1-\frac{1}{2}\left[1-\frac{u_e\, \beta^{-1/2}}{(\Omega/k)}\right]
\end{equation}

\noindent which has been adapted from \cite{sfakiotakisReviewFishSwimming1999} using non-dimensional variables. It was shown in \cite{sfakiotakisReviewFishSwimming1999} that $\eta^*$ is restricted to lie in the range $[0.5, 1.0]$. 

The motion of the flexible propulsor is comprised of four traveling waves - see \eqref{eq18}; therefore the condition in \eqref{eq20} and the expression in \eqref{eq21} are inapplicable. We can however compute a value of the average non-dimensional wavenumber, which we denote by $\bar{k}$; simulation results show that positive thrust is generated when \eqref{eq20} is satisfied with $k$ replaced by $\bar{k}$. To compute $\bar k$, we first recognize that $v(u,\tau)$ of \eqref{eq7} is a complex helix defined by the shape function $f(u)$ which rotates with angular velocity $\Omega$. At any given time, $\bar k$ can be computed from the phase of the helix $\phi(u)$ as follows
\begin{equation}\label{eq22}
\bar k = \phi(0) - \phi(1), \qquad \phi(u) = \atantwo(\operatorname{Im}[f(u)], \operatorname{Re}[f(u)])
\end{equation}

\begin{figure}[t!]
\centering
\psfrag{Q}[][]{\small{$\phi(u)$}}
\psfrag{R}[][]{\small{$u$}}
\psfrag{S}[][]{\small{$(d\phi/du)$}}
\psfrag{C}[][]{\small{slope:$\!$ -$\bar k$}}
\psfrag{B}[][]{\small{$\phi(u)$}}
\psfrag{A}[][]{\small{$(d\phi/du)$}}
\includegraphics[width=0.55\textwidth]{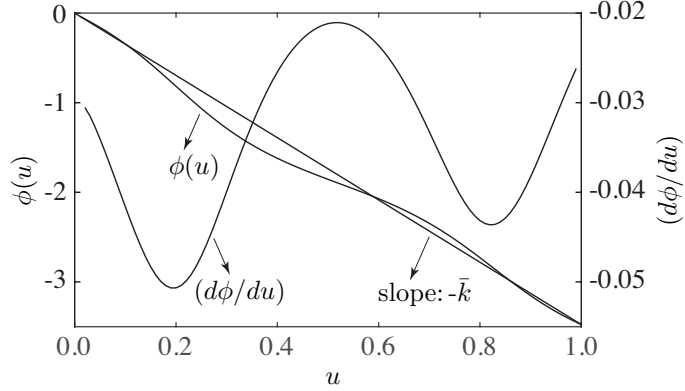}
\caption{\small{Phase plot for Case 6 at $u_e = 3.4$ and $\gamma = 0.9$}}
\label{Fig12}
\end{figure}

For the purpose of illustration, we plot $\phi(u)$ and $\bar k$ for the specific operating point: $u_e = 3.4$ and $\gamma = 0.90$ for Case 6 - see Fig.\ref{Fig12}. It can be seen that $\phi(u)$ decreases as $u$ varies from $0$ to $1$, which signifies that the waveform travels from the hinged end to the free end of the beam. The value of $\bar k$ is the negative of the slope of the straight line joining $\phi(0)$ and $\phi(1)$. The phase angle $\phi$ in Fig.\ref{Fig12} is observed to exhibit undulations about the straight line, implying that the phase does not vary linearly. To characterize this variation, we plot the spatial derivative of the phase angle $(d\phi/du)$ in Fig.\ref{Fig12} and define the phase smoothness factor (PSF):
\begin{equation} \label{eq23}
{\rm PSF} = \frac{\min|(d\phi/du)|}{\max|(d\phi/du)|}
\end{equation}

\noindent The value of PSF signifies the extent to which the traveling wave exhibit a stop-and-go motion as it moves from the hinged end ($u = 0$) to the free end of the flexible tail ($u = 1$). When $\phi(u)$ varies linearly, the value of PSF is equal to 1, which describes a wave traveling with constant velocity. In the next section, it will be shown that the Froude efficiency is highly correlated with the value of PSF, with higher efficiencies associated with PSF values closer to unity.

\subsection{Illustrative Examples of Traveling Waveforms}\label{sec63}
\subsubsection{Effect of Change in $u_{\rm e}$ and $\gamma$}\label{sec631}

We illustrate the change in the traveling waveform and its propulsive characteristics due to changes in the external flow velocity $u_{\rm e}$ and the location of sensing $\gamma$. We consider three operating points from Case 6, presented in Section \ref{sec56}: a nominal point (Point 1) and two other points, obtained by varying either $\gamma$ (Point 2) or $u_{\rm e}$ (Point 3). The critical feedback gain, the critical frequency, the Froude efficiency, the average nondimensional wavenumber, the value of PSF, the wavespeed, and the value of efficiency computed using \eqref{eq21} at these three points are shown in Table \ref{Tab4}.\
\begin{table}[b!]
\centering
\caption{Propulsive characteristics at three operating points of Case 6}
\label{Tab4}
\begin{tabular}{|c|c|c|c|c|c|c|c|c|c|}
\hline
Point & $u_{\rm e}$ & $\gamma$ & $c_{\rm cr}$ & $\Omega = \Omega_{\rm cr}$ & $\eta$ & $\bar k$ & PSF & $(\Omega/\bar{k})$ & $\eta^*$ \\ \hline
1 & 3.2  & 0.30 & 9.78  & 193.1 & 0.514 & 13.40 & 0.074 & 14.41 & 0.611 \\ \hline
2 & 3.2  & 0.90 & 3.65  & 24.0  & 0.701 & 3.24  & 0.403 & 7.40  & 0.716 \\ \hline
3 & 12.4 & 0.30 & 11.25 & 258.1 & 0.751 & 11.11 & 0.505 & 23.23 & 0.767 \\ \hline
\end{tabular}
\end{table}

\begin{figure}[t!]
\centering
\psfrag{E}[][]{\small{$\Omega\tau$}}
\psfrag{D}[][]{\small{$2\pi$}}
\psfrag{C}[][]{\small{$\pi$}}
\psfrag{B}[][]{\small{$0$}}
\psfrag{A}[][]{\small{$u$}}
\includegraphics[width=1\textwidth]{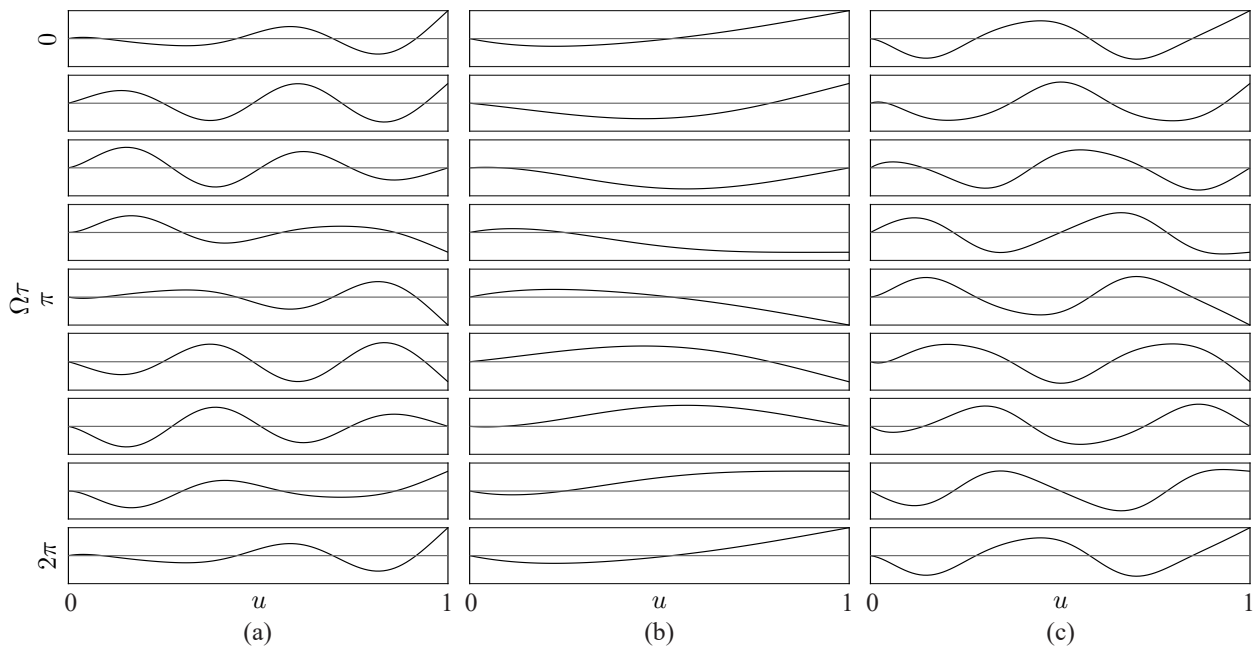}
\caption{\small{Traveling waveforms over one complete cycle, shown at intervals of $\pi/4$, for the three operating points of Case 6 shown in Table \ref{Tab4}: (a) Point 1, (b) Point 2, and (c) Point 3.}}
\label{Fig13}
\end{figure}

For the nominal point (Point 1: $u_{\rm e} = 3.2$, $\gamma = 0.30$), the waveform is shown in Fig.\ref{Fig13} (a). The waveform exhibits a ``stop-and-go" motion with amplitudes varying considerably along the length of the tail, which corresponds to its low PSF value. The value of $\eta^*$ lies near the lower bound of the range $[0.5, 1.0]$; this follows from \eqref{eq21} since the wavespeed $(\Omega/\bar{k})$ is significantly higher than the external flow velocity $u_{\rm e}$\footnote{For our simulations, we assumed $\beta = 0.9822$ - see Section \ref{sec43}. This gives a value of $\beta^{-1/2} = 1.0090 \approx 1$.}. The value of $\eta$ computed using \eqref{eq19} is also near the lower bound of $[0.5, 1.0]$.\

When the location of sensing alone is changed (Point 2: $u_{\rm e} = 3.2$, $\gamma = 0.90$), we notice a significant drop in the frequency with a less significant drop in the wavenumber, which results in a lower wavespeed - see Table \ref{Tab4}. While still producing positive thrust as per \eqref{eq20}, the drop in wavespeed results in a jump in the value of $\eta^*$, which closely matches the computed value of $\eta$. The waveform, shown in  Fig.\ref{Fig13} (b), indicates a fewer number of undulations compared to that of Fig.\ref{Fig13} (a) in accordance with the the lower wavenumber. Additionally, it can be seen that the waveform travels much more smoothly along the length of the tail which is captured by its much higher value of PSF.\

When the external flow velocity alone is changed (Point 3: $u_{\rm e} = 12.4$, $\gamma = 0.30$), we notice a moderate increase in the frequency but a slight decrease in the wavenumber, which results in a higher wavespeed - see Table \ref{Tab4}. Though the wavespeed is increased, the system operates at a much higher external velocity which results in a jump in the value of $\eta^*$, which, again, closely matches the computed value of $\eta$. The waveform, shown in  Fig.\ref{Fig13} (c), indicates a similar number of undulations compared to that of Fig.\ref{Fig13} (a) due to the similar magnitude of the wavenumber. While there are many undulations, it can be seen that the waveform travels much more smoothly along the length of the tail, which is captured by its much higher value of PSF.\

\begin{remark}
The results in Table \ref{Tab4} indicate that there is a strong positive correlation between the value of PSF and the values of $\eta$ and $\eta^*$. This trend has been observed for other points in the $\gamma$-$u_{\rm e}$ domain for Case 6, as well as other cases with different modes of sensing and actuation.
\end{remark}

Video animations of the waveforms for Points 1, 2, and 3 have been uploaded as supplementary material - see Video1.mp4. To illustrate the ``stop-and-go" motion of the waveform for Point 1 and the smooth waveforms of Points 2 and 3, they are presented at the same frequency.\

\subsubsection{Dependence of Propulsive Characteristics on $u_{\rm e}$ and $\Omega$}\label{sec632}
\begin{table}[t!]
\centering
\caption{Propulsive characteristics at three operating points of Cases 1, 2, 4 with identical $u_e$ and similar $\Omega$ values}
\label{Tab5}
\vspace{0.05in}
\begin{tabular}{|c|c|c|c|c|c|c|c|c|c|c|}
\hline &&&&&&&&&\\[-2.25ex]
Case & $u_{\rm e}$ & $\gamma$ & $c_{\rm cr}$ & $\Omega = \Omega_{\rm cr}$ & $\eta$ & $\bar k$ & PSF & $(\Omega/\bar{k})$ & $\eta^*$ \\ \hline
1 & 6.9 & 0.30 & 1.46 & 152.8 & 0.720 & 9.51 & 0.318 & 16.073 & 0.715 \\ \hline
2 & 6.9 & 0.11 & 0.11 & 152.2 & 0.723 & 9.47 & 0.318 & 16.074 & 0.715 \\ \hline
4 & 6.9 & 0.92 & 1.24 & 145.7 & 0.746 & 9.06 & 0.324 & 16.079 & 0.715 \\ \hline
\end{tabular}
\end{table}

\begin{figure}[b!]
\centering
\psfrag{A}[][]{\small{$u$}}
\psfrag{B}[][]{\small{$\Omega\tau = 0$}}
\psfrag{Q}[][]{\small{$0$}}
\psfrag{R}[][]{\small{$1$}}
\includegraphics[width=0.55\textwidth]{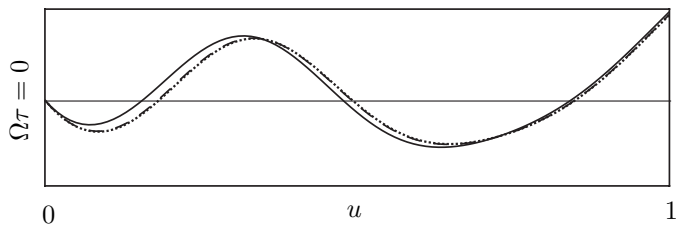}
\caption{\small{Traveling waveforms for the three operating points of Case 1 (dotted),  Case 2 (dashed), and Case 4 (solid), shown in Table \ref{Tab5}. The waveforms from Case 1 and 2 exhibit tremendous overlap while the waveform from Case 4 is slightly separated.}}
\label{Fig14}
\end{figure}

We select three operating points with identical values of $u_{\rm e}$ and similar values of $\Omega$\footnote{For a fixed value of $u_{\rm e}$, each critical frequency plot degenerates to a line where each point on the line corresponds to a different value of $\Omega_{\rm cr}$. Through trial and error it is possible to find similar $\Omega_{\rm cr}$ values across multiple cases.} from three different cases, namely Case 1, Case 2, and Case 4. The propulsive characteristics of these three cases are found to be very similar - see Table \ref{Tab5}. These characteristics, which include the efficiency, the average nondimensional wavenumber, the value of PSF, and the wavespeed, are particularly close for Cases 1 and 2, for which the frequencies differ by only $0.4\%$. The characteristics of Case 4 differ slightly more as its frequency is $4.5\%$ lower than the other two cases. It is observed that for a constant $u_e$, a lower frequency is associated with higher values of $\eta$ and PSF; this agrees well with the trend illustrated by Points 1 and 2 in Section \ref{sec631}. The waveforms for the three cases are shown in Fig.\ref{Fig14}. The waveforms for Cases 1 and 2 are nearly indistinguishable, while the waveform for Case 4 is slightly different from the other two, which can be attributed to the slightly different value of the frequency of oscillation.\

\begin{remark}
The data in Table \ref{Tab5} are a small set of results which indicate that the nature of a waveform and its propulsive characteristics depend solely on the values of $u_{\rm e}$ and $\Omega$, and are independent of the modes of actuation and sensing.
\end{remark}

Video animations of the waveforms for the specific operating points of Cases 1, 2, and 4 have been uploaded as supplementary material - see Video2.mp4. The waveforms are barely distinguishable; this illustrates the dependence of the waveform solely on the external flow velocity and the critical frequency, and not on the modes of actuation and sensing or the location of sensing.\

The results observed in this section, that the waveforms are uniquely determined by the values of $u_{\rm e}$ and $\Omega_{\rm cr}$ and are independent of the modes of sensing and actuation, can be explained mathematically as follows. For convenience, we revisit \eqref{eq16}, \eqref{eq11}, and \eqref{eq13} from Sections \ref{sec3} and \ref{sec4}:
\begin{equation*}
v(u, \tau) = \sum_{n=1}^{4} A_n\, e^{{\rm Re}[z_n]u}\, e^{i \left\{{\rm Im}[z_n]u + {\rm Re}[\Omega_{\rm cr}]\tau\right\}}
\tag{\ref{eq16} revisited}
\end{equation*}

\begin{equation*}
z^4 + u_{\rm e}^2 z^2 + 2 u_{\rm e} \sqrt{\beta}\, i \Omega z - \Omega^2 = 0
\tag{\ref{eq11} revisited}
\end{equation*}

\begin{equation*}
\underbrace{\begin{bmatrix}
1 & 1 & 1 & 1 \\
z_1^2 e^{z_1}  & z_2^2 e^{z_2}  & z_3^2 e^{z_3}  & z_4^2 e^{z_4}  \\
z_1^3 e^{z_1} & z_2^3 e^{z_2}  & z_3^3 e^{z_3}  & z_4^3 e^{z_4} \\
\delta_1 & \delta_2 & \delta_3 & \delta_4
\end{bmatrix}
}_{\mathbb{Z}}
\begin{bmatrix} A_1 \\ A_2 \\ A_3 \\ A_4 \end{bmatrix} =
\begin{bmatrix} 0 \\ 0 \\ 0 \\ 0 \end{bmatrix} 
\tag{\ref{eq13} revisited}
\end{equation*}

\noindent The waveform in \eqref{eq16} is defined by three terms: $\Omega_{\rm cr}$, $z_n$, and $A_n$, of which $\Omega_{\rm cr}$ is the same for the different cases. From \eqref{eq11}, we can see that for a fixed value of $\beta$, the solutions of $z_n$, $n = 1, 2, 3, 4$, are uniquely defined by $u_{\rm e}$ and $\Omega_{\rm cr}$ and are independent of the modes of sensing and feedback. As a consequence, the first three rows of $\mathbb{Z}$ in \eqref{eq13} are solely functions of $u_{\rm e}$ and $\Omega_{\rm cr}$. It can be shown that the first three rows of $\mathbb{Z}$ are linearly independent though $\mathbb{Z}$ is singular as that is how $\Omega_{\rm cr}$ was solved for. Therefore, the direction of the null space vector $(A_1,A_2,A_3,A_4)^T$ is independent of the last row of $\mathbb{Z}$ and thus independent of the modes of sensing and feedback. Therefore, each term of the waveform in \eqref{eq16} is solely a function of $u_{\rm e}$ and $\Omega_{\rm cr}$.\

\section{Conclusion}\label{sec7}

Beam flutter occurs due to non-conservative loading. Such loading is commonly generated by a follower force or through interaction of the beam with a fluid, flowing internally or externally. It is also possible to produce non-conservative loading by applying an actuation, proportional to some state of the beam, at one of the boundaries. This work provides a generalized description of flutter, generated using this method, in a pinned-free beam. The actuation can take the form of a moment or an angle prescribed at the pinned boundary. This actuation is proportional to some state of the beam (displacement, slope, or curvature) measured at any location along its length. The onset of flutter is not symmetric about zero gain, meaning that there are twelve potential flutter mechanisms identified here: two modes of actuation, three modes of sensing, and two signs of the critical gain for each combination. This opens up a wide range of physical mechanisms, beyond the standard follower force and fluid-flow mechanisms, that can be used to produce and study flutter, some of which we intend to realize in future work.\

For each combination of actuation and sensing, the critical gain was determined over a range of external flow velocities and sensing locations along the beam. For a majority of the twelve possible combinations, stability was lost through flutter; a representative sample of six cases (one for each combination of actuation and sensing) were investigated. These six cases illustrated a rich set of stability transitions that depend strongly on the location of sensing and mildly on the external flow velocity. It was observed that small changes in the location of sensing could result in very different modes of flutter with large jumps in the critical frequency thereby resulting in significantly different traveling waveforms.\

These traveling waveforms of the flexible beam could be exploited to develop a propulsion mechanism for underwater vehicles. Because the Euler-Bernoulli beam model is fourth-order in space, the solution naturally comprises four separate traveling waves. Based on the spatial derivative of the beam phase, we constructed a metric, ``the phase smoothness factor", which is a measure of how closely the four traveling waves can be approximated by a single waveform. Waveforms which are smoother demonstrate higher propulsive efficiency. The same is observed in nature, where fish swim with optimized waveforms that are smooth and efficient. Interestingly, the propulsive characteristics of the beam do not depend on the combination of actuation and sensing by which flutter is produced; they depend only on the values of the dimensionless fluid velocity and critical frequency, which completely define the waveform.\

In addition to experimental validation of feedback-induced flutter towards generating a traveling wave, we envision a number of other directions along which this work can be extended. First, the observation that a waveform dominated by a single traveling wave appears to be more efficient than a mixed waveform is satisfying, but a proof has eluded us to date. It is also likely that analyzing a discrete analogy of the problem, in which flutter of an articulated system is produced through actuation of the base, can provide additional insights because of the reduction to finite dimension. Finally, we believe that the results can be extended to an interesting boundary condition at the free end, that of a hydrodynamically ``active’’ fin, which has mass, dimension, and experiences fluid forces. This conceptually mimics the situation of a fast-swimming thunniform fish, which has a large caudal fin at the end of its tail.

\bibliography{References.bib}
\bibliographystyle{elsarticle-num}

\end{document}